\documentclass[reqno,11pt]{article}
\usepackage{amssymb,amsthm,amsmath}
\usepackage{epsfig}

\theoremstyle{plain}
\begingroup
\newtheorem*{metathm}{Metatheorem} 
\newtheorem{thm}{Theorem}[section] 

\newtheorem{lemma}[thm]{Lemma}

\endgroup

\newcommand{\nwc}{\newcommand}
\nwc{\Levy}{L\'evy}
\nwc{\qref}[1]{(\ref{#1})}
\nwc{\D}{\partial}
\nwc{\mmt}{m}
\nwc{\fzero}{F_\rho} 
\nwc{\fone}{M_\rho} 
\nwc{\ip}[1]{\langle #1 \rangle}
\nwc{\ipbig}[1]{\left\langle #1 \right\rangle}
\nwc{\Lip}{\mathop{\rm Lip}\nolimits}
\nwc{\Tgel}{T_{\rm gel}}

\renewcommand{\Im}{\mathop{\rm Im}\nolimits}

\theoremstyle{definition}
\newtheorem{defn}[thm]{Definition} 
\newtheorem{rem}[thm]{Remark}

\theoremstyle{remark}
 
\numberwithin{equation}{section}
\numberwithin{figure}{section}

\begin{document}
\title{Approach to self-similarity in
  Smoluchowski's coagulation equations} 
\author{Govind Menon\textsuperscript{1} and Robert. L.
Pego\textsuperscript{2}}

\date{June 2003}

\maketitle

\begin{abstract}
We consider the approach to self-similarity (or dynamical scaling) in
Smoluchowski's equations of 
coagulation for the solvable kernels $K(x,y)=2,x+y$ and $xy$.
In addition to the known self-similar solutions with exponential tails, 
there are one-parameter families of solutions with algebraic decay, 
whose form is related to heavy-tailed distributions
well-known in probability theory. 
For $K=2$ the size distribution is Mittag-Leffler, and for
$K=x+y$ and $K=xy$ it is a power-law rescaling of a maximally skewed
$\alpha$-stable \Levy\ distribution.
We characterize completely the domains of attraction of all
self-similar solutions under weak convergence of measures. Our results
are analogous to the classical characterization of stable distributions in
probability theory. The proofs are simple, relying on the Laplace
transform and a fundamental rigidity lemma for scaling limits.
\end{abstract}

\bigskip

\footnotetext[1]
{Department of Mathematics, University of Wisconsin, Madison WI 53706.
Email: menon@math.wisc.edu}
\footnotetext[2]{Department of Mathematics \&
Institute for Physical Science and Technology,
University of Maryland, College Park MD 20742. 
Email: rlp@math.umd.edu}

\pagebreak
 \section{Introduction}
Smoluchowski's coagulation equations provide a mean field description
of several processes of mass aggregation in nature.  We study the
evolution of $n(t,x)$, the number of clusters of mass $x$ per unit
volume at time $t$. Clusters of mass $x$ and $y$ coalesce by binary collisions 
at a rate governed by a symmetric kernel $K(x,y)$, whence
\begin{equation}
\label{eq:smol1}
\frac{\partial n}{\partial t}(t,x) = \frac{1}{2} \int_0^x K(x-y,y) n(t,x-y)
n(t,y) dy  - \int_0^\infty K(x,y) n(t,x) n(t,y) dy. 
\end{equation}
All microscopic interactions are subsumed into
the agglomeration rate kernel $K$, 
and the process is assumed to be stationary in
space. A broad survey of applications, especially in physical
chemistry,  may be found in the article by Drake~\cite{Drake}.
Equation (\ref{eq:smol1}) has been used in an amazingly diverse range
of applications, such as the formation of clouds and
smog~\cite{Friedlander},  the clustering of planets, stars and
galaxies~\cite{Silk}, the kinetics of polymerization~\cite{Ziff}, and
even the schooling of fishes~\cite{Niwa} and the formation of ``marine
snow'' (see \cite{Kiorboe}). In the past few years, there
has been a resurgence of mathematical interest in the field, largely
due to the work of probabilists. An influential survey article by
Aldous summarizes the recent state of affairs~\cite{Aldous}.

An issue of importance for homogeneous kernels, 
kernels that satisfy
$K(\alpha x, \alpha y) = \alpha^\gamma K(x,y)$, is the phenomenon of
 {\em dynamical   scaling\/} for all initial data in a {\em
 universality class\/}.  Mathematically, this corresponds
  to the problem of existence  of {\em scaling} 
or {\em self-similar solutions} and characterization of their {\em domains
  of attraction\/}.  In the aerosols community the relevant rubric
is the theory of self-preserving spectra, which is treated at length
in Friedlander's book~\cite{Friedlander} and the extensive survey
of Drake~\cite{Drake}.  
For a large class of kernels with $\gamma \leq 1$, there is
numerical evidence that 
solutions evolve to a self-similar form~\cite{Lee}. There
are also physical self-consistency arguments  that have been used to
derive asymptotics for scaling  solutions~\cite{DE1}. 
In the case $\gamma >1$, for a general class of kernels 
it is known that solutions must lose mass
(presumably to infinite-mass clusters) 
in finite time~\cite{Escobedo,Jeon}, but there is no general 
rigorous result on the precise nature of this blow-up in mass
transport. In several instances the known solutions 
have unphysical divergences such as infinite mass. Thus, a general
existence theory for finite-mass self-similar solutions, for example, 
would be of some value. 

The kernels $K(x,y) = 2, x+y$ and $xy$ play a
special role, as (\ref{eq:smol1}) can then be solved by the
Laplace transform. It is widely known that each of these kernels
admits a self-similar solution with exponential decay (see Table 2 in
~\cite{Aldous}). These kernels are also special since certain solutions to
  (\ref{eq:smol1}) can be viewed as ergodic averages in beautiful
probabilistic constructions, involving thinning of
  renewal processes ($K=2$), and tree-valued Markov processes and
  their self-similar limits ($K=x+y, xy$)~\cite{Aldous2}. 
  The additive kernel also figures in interesting recent applications
  given by Bertoin. It provides a natural
  probabilistic interpretation of a sticky particle model related to
  Zeldovich's model of gravitational clustering~\cite{Bertoin2}.
  Also, the known self-similar solution 
  appears in a simple model of turbulence, the inviscid Burgers
  equation with Brownian-motion initial data,
  as the characteristic measure for a Poisson point process
  which describes the shock strengths~\cite{Bertoin3}.

In short, aside from heuristics and numerics,  there are no rigorous
mathematical proofs of the  
existence of self-similar solutions and the approach to self-similar
form for general kernels (see~\cite[Sec 2]{Aldous}). 
And there are only a few partial results for the
solvable kernels:  For $K=2$, Kreer and Penrose~\cite{KP} proved 
local uniform convergence to the scaling solution under some technical
hypotheses on initial data. 
(Also see \cite{daCosta} regarding the discrete case.)
A simple weak convergence theorem in this case follows
from a classical result on the thinning of renewal
processes~\cite{Aldous}. In a recent article, Deaconu and Tanr\'{e}
proved a weak convergence result for all three kernels, but under 
restrictive hypotheses on initial data~\cite{DT}.
Aldous and Pitman have studied the ``eternal additive coalescent,''
and  Bertoin has  characterized ``eternal solutions'' to the Smoluchowski
equation with additive kernel, solutions defined globally for 
$-\infty<t<\infty$~\cite{Aldous-Pitman,Bertoin}. 
Bertoin showed these solutions correspond to the \Levy\ measure of a
first-passage process related to \Levy\ processes with no positive
jumps, and as a particular consequence he derived a new family of self-similar
solutions related to the \Levy\ stable laws of probability.  

In this article we find new families of self-similar solutions for the
constant kernel, rederive the self-similar solutions to the
additive and multiplicative kernels by analytical means, and
characterize all possible domains of attraction under weak convergence
for all the solvable kernels.  We show that  

\begin{enumerate}
\item For each of the solvable kernels, Smoluchowski's equation
admits a one-parameter family of
  scaling solutions parametrized by a number $\rho\in(0,1]$ that
  characterizes the rate of divergence of the $(\gamma+1)$-{st} moment
  of the number density. For $\rho=1$ these solutions reduce to the
  known solutions with exponential tails, while for $0<\rho<1$ the
  number density has algebraic decay (``fat tails'').
  For $K=2$ ($\gamma=0$) the normalized size distribution is a Mittag-Leffler
  distribution as studied by Pillai~\cite{Pillai}.
  For $K=x+y$ ($\gamma=1$) and $xy$ ($\gamma=2$) 
  the $\gamma$-th moment distributions are
  transformed by power-law rescaling to the {\Levy\ stable laws} of 
  probability theory (see~\ref{eq:plaw} and \cite{Bertoin}). 

\item The domains of attraction (under weak convergence of measures) for 
  any scaling solution is determined by a condition on the tails of
  the initial data --- the algebraic rate of divergence of the 
  $(\gamma+1)$-st moment.  A precise characterization is given via
  Karamata's notion of regular variation. In particular, 
  with suitably diverging $(\gamma+1)$-st moment there are
  initial data for which there is no convergence to any self-similar solution.
\end{enumerate}
The self-similar solutions can all be captured by 
expressing their $\gamma$-th moment distribution in the general form
\begin{equation}
x^\gamma n(t,x) = \mmt_\gamma(t) \lambda_\gamma(t)^{-1}
f_{\rho,\gamma}(x\lambda_\gamma(t)^{-1}),
\end{equation}
where explicitly, with $\beta=\rho/(1+\rho)$, 
\begin{eqnarray}
& m_0(t)=t^{-1},\quad  m_1(t)=1,\quad m_2(t)=(1-t)^{-1}, &\\
& \lambda_0(t)=t^{1/\rho},\quad \lambda_1(t)=e^{t/\beta}, \quad
\lambda_2(t)=(1-t)^{-1/\beta}, &
\end{eqnarray}
and the $f_{\rho,\gamma}$ are probability densities given by
\begin{eqnarray}
\label{eq:f_rho_const}
f_{\rho,0}(x) &=& 
\sum_{k=1}^\infty \frac{ (-1)^{k-1}x^{\rho k-1}}{\Gamma(\rho k)},
\\
\label{eq:f_rho_add}
f_{\rho,1}(x)= f_{\rho,2}(x) &=& 
\frac1\pi \sum_{k=1}^\infty
\frac{(-1)^{k-1}x^{k\beta-1}}{k!}\Gamma(1+k-k\beta)\sin
k\pi\beta.\quad
\end{eqnarray}

We work with measure valued solutions to (\ref{eq:smol1}) denoted
$\nu_t$, where $\nu_t((a,b])$ denotes the number of clusters with size
$x\in(a,b]$: $\nu_t((a,b])=\int_a^b n(t,x)\,dx$ if 
$\nu_t$ has integrable density $n(t,x)$.
For each of the solvable kernels, we  associate a natural probability
distribution function $F(t,x)$ to the solution:
\begin{equation}
F(t,x) = \int_0^x y^{\gamma}\nu_t(dy)\left\slash
\int_0^\infty y^{\gamma}\nu_t(dy)\right. .
\end{equation}
This is the
size-biased  distribution for $K=2$, the mass distribution for $K=x+y$ and
the second  moment distribution for $K=xy$. We are interested in
necessary and sufficient conditions for the
convergence of a rescaling $F(t, \lambda(t)x)$ to a nontrivial 
limit $F_*(x)$. Our results may be summarized in the following.

\begin{metathm}
\label{mthm:main}
For the kernels $K(x,y)=2$, $x+y$, $xy$ with degree of
homogeneity $\gamma=0$, 1, 2 respectively,
let $T_\gamma=\infty$ for  $\gamma =0, 1$ and $T_\gamma =\Tgel$ for
$\gamma=2$.  Then for any solution of Smoluchowski's coagulation
equation, there is a rescaling 
$\lambda(t)$ and a nontrivial probability distribution function 
$F_*$ such that
\[
{ \lim_{t \rightarrow T_\gamma} F(t, \lambda(t)x)
 =F_*(x)\;\mbox{at all points of continuity of $F_*$}}
\]
if and only if
\[
\int_0^x y^{\gamma+1} \nu_0(dy) \sim x^{1-\rho}
L(x) \quad\mbox{as $x \rightarrow \infty$,} 
\]
where $\rho \in (0,1]$ and $L(x)$ is a
function slowly varying at  infinity. In the converse implication,
$F_*$ must be   a rescaling of
$F_{\rho,\gamma}(x)=\int_0^xf_{\rho,\gamma}(y)\,dy$.
\end{metathm}

Precise statements are deferred to
Theorem~\ref{thm:const}, Theorem \ref{thm:add}, and
Theorem~\ref{thm:mult}. We write the results in this form to stress
the analogy with the classical  
characterization of the \Levy\ stable distributions in probability
theory~\cite{Feller}. For the additive and
multiplicative kernels the analogy is an intimate relation with
distributions for asymmetric \Levy\ flights: 
{\em the self-similar solutions can be transformed by a power-law rescaling
into maximally skewed $\alpha$-stable \Levy\ distributions\/}.
A deeper understanding of our results is obtained from 
Bertoin's  study of eternal
solutions~\cite{Bertoin}. The eternal solutions are
analogous to infinitely divisible distributions of probability
theory. Let $\nu_t$ be the value of an eternal
solution at time $t$. Then, loosely speaking, for any $s <t$ the  
measure $\nu_s$ decomposes $\nu_t$ such that  
$\nu_t$ is reconstituted from  $\nu_s$ under coagulation. This
heuristic statement is made precise by Bertoin's  characterization of
\Levy\ pairs  $(\sigma^2, \Lambda)$ for the eternal solutions. This  is  the
analogue of the classical L\'{e}vy-Khintchine characterization of the
infinitely divisible distributions~\cite{Feller}. Among the infinitely
divisible distributions, the stable distributions are of special interest, and
their \Levy\ canonical measures are pure power laws. 
And indeed, the self-similar solutions to Smoluchowski's equation
(\ref{eq:smol1}) have \Levy\ pairs corresponding to pure power laws:
$\sigma^2=0, \Lambda(dx) = cx^{-\alpha-1} dx, 1< \alpha <2$ and 
$\sigma^2=1$ and $\Lambda=0$ for $\alpha=2$ exactly as in the
classical  characterization. When viewed in this context, our theorems
are entirely natural. 

For $K=2$ the main theorem may be interpreted probabilistically as a stability
result for renewal processes on the line under uniform thinning
(see~\cite{MPY}).  For $K=x+y$ the results are related to Burgers
turbulence, for solutions of the inviscid Burgers' equation when the initial
velocity is given by a \Levy\ process with no positive
jumps~\cite{Bertoin3}. 


We exploit the analogy with the classical limit theorems of
probability to obtain simple proofs of optimal theorems. The proofs
involve little more than the solution for the Laplace
transform and a fundamental rigidity lemma that characterizes scaling
limits via functions of regular variation~\cite[VIII.8.3]{Feller}. 
And the analogy extends much further. The central limit theorem is perhaps
the most intensively studied result in probability theory. Thus, 
we can demand stronger forms of convergence as in expansions
related to the central limit theorem. 
In companion articles we plan to study (a) uniform convergence of densities 
to the self-similar solutions with exponential tails (in analogy
with the uniform convergence of 
densities in the central limit theorem)~\cite{MP2},  
(b) metric estimates (in analogy with the 
Berry-Ess\'{e}en theorem) and (c) large deviation
estimates. We have found proofs for (a), and  
partial results for (b) and (c), that follow easily from
a combination of the solution formula and the classical method of
characteristic functions outlined in Feller~\cite{Feller}.

It is worth remarking that the folklore in the applied literature is that the
scaling  solutions  are unique (e.g., see~\cite{DE1,Friedlander}). 
This is false in general (though the solutions with
exponential tails are indeed special --- they attract all solutions
with finite $(\gamma+1)$-st moment).  With
hindsight, this nonuniqueness is 
not surprising. The existence of a one-parameter family of scaling
solutions is well known in physically related mean field 
models which show coarsening, such 
as the Lifshitz-Slyozov-Wagner  model~\cite{NP1}, one-dimensional
models for the coalescence of droplets~\cite{Derrida}, and in 
cut-and-paste models of coarsening~\cite{Gallay}. From the
mathematical point of view the fundamental role of regular variation
in branching processes is well 
established~\cite{Bingham} and it is only natural that it
should reappear in the  ``dual process'' of coalescence. 
We conjecture that analogous results hold for
general homogeneous kernels, but these lie beyond our
techniques based on the Laplace transform.

\nwc{\num}{\nu}  
\nwc{\mx}{\min\{1,x\}}
\nwc{\my}{\min\{1,y\}}
\section{Well-posedness for measure valued solutions}
\subsection{Desingularized Laplace transforms}
\label{sec:renorm}
Smoluchowski's equations determine a process of mass transport, and
measure valued solutions are an appropriate mathematical abstraction
that contains solutions to the discrete and continuous coagulation
equations within a unified framework.
Norris recently proved several strong results for
well-posedness of measure valued solutions, but these do not 
apply with quite the generality 
we prefer for $K=x+y$ and $K=xy$~\cite{Norris}. 
We work with a somewhat different notion of solution motivated
by the explicit solution obtained with the Laplace transform. The use
of the Laplace transform for these kernels is classical~\cite{Drake},
and aside from trivial changes of notation, many of the equations below
may be found in Bertoin's article~\cite{Bertoin}. 
While our primary aim in writing this article is not to tackle the
question of well-posedness, we show below that 
the divergence of the self-similar
solutions necessitates some care in the definition of solutions.
The impatient reader may skim through this section making a note of
the main theorems and the explicit solution formulas.
Our primary source for background on the Laplace transform
is Feller~\cite{Feller}.

Let $E$ denote the open interval $(0,\infty)$ and let $\mathcal{M}_+$
denote the space of positive Radon measures on 
$(0, \infty)$. 
We interpret the number of clusters of size
$x \in(a,b]$ per unit volume as $\num((a,b])$ for $\num \in
\mathcal{M}_+$. We use the same letter to denote the 
distribution function of the measure, writing $\num(x)=\num((0,x])$,
if this quantity is finite. We let $\mmt_p=\int_Ex^p\nu(dx)$
denote the $p$-th moment of the measure, so $\mmt_0$ is the total
number of clusters and $\mmt_1$ is the total mass.

We let $\eta(s)$ be the Laplace transform
of $\num(x)$ defined by  
\begin{equation}
\nonumber
\eta(s) = \int_E e^{-sx} \num(dx) = \int_E e^{-sx} n(x) dx.
\end{equation}
The last equation holds when $\num$ has a density $n$. 
In what follows we need to work with time-dependent measures $\nu_t$ 
for which the total number of clusters and/or total mass may be infinite. 
Consequently, it is more convenient to work with the variables (the
``desingularized Laplace transform'') given by 
\begin{equation}
\label{eq:renorm_6}
\varphi(t,s) = \int_E (1-e^{-sx}) \nu_t(dx), 
\quad 
\psi(t,s) = \int_E (1-e^{-sx}) x \,\nu_t(dx).
\end{equation}
The variable $u=\D_s\varphi$ has the 
important physical interpretation that it is
the Laplace transform of the mass measure. Probabilists will recognize
the obvious similarity to the L\'{e}vy-Khinchine representation.   
The equations of evolution in terms of these variables
are extremely simple: 

\begin{eqnarray}
\label{eq:evol-const}
&\D_t \varphi = -\varphi^2  &
\qquad\mbox{for $K=2$}, \\
\label{eq:evol-add}
&\D_t \varphi -\varphi \D_s\varphi = -\varphi &
\qquad\mbox{for $K=x+y$}, \\
\label{eq:mult3}
&\partial_t \psi -\psi \partial_s \psi =0 &
\qquad\mbox{for $K=xy$.}
\end{eqnarray}
We will construct measures using these equations and establish
that they are solutions of Smoluchowski's equation in an appropriate weak sense.

The function $1-e^{-sx}$ does not have compact support in $E$, but has
finite limits at $0$ and $\infty$. A simple way to treat these limits
is to consider $\bar{E}=[0,\infty]$, and consider continuous functions
on $\bar{E}$, where $f(\infty)$ always means $\lim_{x \to\infty} f(x)$. 
\begin{defn}
\label{defn:spaces}
\begin{enumerate}
\item $C(\bar{E})$ is the space of continuous maps $f:\bar{E} \rightarrow
  \mathbb{R}$ equipped with the norm $\|f\|_{C(\bar{E})} = \sup_x |f(x)|$. 
\item $C^k(\bar{E})$ consists of $k$ times continuously differentiable
  functions on $E$ such that $f,\ldots, f^{(k)} \in C(\bar{E})$. It is equipped
  with the norm  
$\|f\|_{C^k(\bar{E})} = \|f\|_{C(\bar{E})} +\ldots+
  \|f^{(k)}\|_{C(\bar{E})}$. 
\item $C^k_c(E)$ is the subspace of $C^k(E)$ with compact
  support in $E$.
\item $\mathcal{E}_k$ denotes the subspace of $C^k(\bar{E})$ of
  functions 
  whose derivatives up to order $k$ decay exponentially, that is 
$\sum_{j=1}^k |f^{(j)}(x)| \leq C_f e^{-\alpha x}$ for some $\alpha
>0$,
  and 
  whose derivatives up to order $k-1$ vanish at zero, that is
  $f^{(j)}(0)=0$ for $j<k$.
\end{enumerate}
\end{defn}
The following classical approximation lemma shows that the functions
$1-e^{-sx}$ span a dense set in $C(\bar{E})$.
\begin{lemma}
\label{le:bernstein}
\begin{enumerate}
\item[(a)] Let $f \in C(\bar{E})$. Then for every $s>0$ there is a sequence   
$P_n(x;s) = \sum_{k=1}^n a_{n,k}(s) (1-e^{-sx})^k$ such that 
\[\lim_{n \to \infty} \|P_n(x;s) -f(x)\|_{C(\bar{E})} =0.\]
\item[(b)] If $f \in \mathcal{E}_k$ then we also have
$\lim_{n \to \infty} \|P_n(x;s) -f(x)\|_{C^k(\bar{E})} = 0$ for
sufficiently small $s$.
\end{enumerate}
\end{lemma}
\begin{proof}
The problem may be reduced to polynomial approximation on
the unit interval by the transformation $y = 1-e^{-sx}$, and $g(y)=f(x)$. 
Then $g \in C([0,1])$ for  $f \in C(\bar{E})$. Assertion (a) now follows from
Weierstrass' approximation theorem. A particularly useful choice 
are the Bernstein polynomials $B_{n,g}(y)$ of $g$, and
$P_n(x;s)= B_{n,g}(y)$. 
Suppose $g \in C^k[0,1]$. It is then classical that 
$\lim_{n \to \infty} \|g(y)-B_{n,g}(y)\|_{C^k[0,1]}
=0$~\cite[p.25]{Lorentz}. Thus, in order to obtain (b) it suffices to
show that $g \in C^k[0,1]$. A decay assumption on $f$ is warranted,
because by the chain rule
\[ g'(y) = f'(x)\frac{e^{sx}}{s}, \quad g''(y) =  \frac{e^{2sx}}{s^2}
\left (f''(x)  + s f'(x)
\right), \quad\mbox{etc.}\] 
But $f \in \mathcal{E}_k$, so that  $\sum_{j=1}^k |f^{(j)}(x)| \leq
C_f e^{-\alpha x}$. Hence, for $s<\alpha/k$ we have   
$\lim_{y \to 1} g'(y)= \ldots =\lim_{y \to 1} g^{(k)}(y)=0$ and 
$g \in C^k[0,1]$. Thus, given any $\varepsilon >0$, there is an
$n(\varepsilon)$ such that $|g'(y)-B_{n,g}'(y)|< \varepsilon$.
The change of variables now works to our advantage, for
we have $|f'(x) - P_n'(x)| < \varepsilon se^{-sx}$. 
A similar calculation holds for all $k$ derivatives proving (b).
\end{proof}

\subsection{A weak formulation for measure valued solutions}
Following Norris, we will generalize Smoluchowski's equation as
follows. To every {\em finite\/}, positive measure $\nu$ we associate the
measure $L(\nu)$ defined by duality with continuous functions with
compact support. 
\begin{equation}
\label{eq:defn_L}
\ip{f,L(\nu)} = \frac{1}{2} \int_E\int_E [f(x+y)-f(x)-f(y)]
K(x,y) \,\nu(dx) \nu(dy).
\end{equation}
It is then natural to consider the weak formulation 
\begin{equation}
\label{eq:weak1}
 \ip{f,\nu_t} = \ip{f,\nu_0} + \int_0^t \ip{f, L(\nu_\tau)}\, d\tau, \quad \mbox{for
 every} \;\;f \in C_c(E).
\end{equation}
This suffices for the case $K=1$, but it is insufficient
for $K=x+y$ and $K=xy$. The self-similar solutions to these 
kernels are not finite measures, and consequently they are not solutions in
the sense of Norris, since they fail condition (3) in his
definition~\cite[p.80]{Norris}.

The basic obstruction is that
$L(\nu_t)$ is not a measure in general, since  
$\ip{f, L(\nu_t)}$ may not be finite for all continuous functions. 
The reason is that even
though $f$ may have compact support in $E$ the function
$Tf(x,y) := f(x+y)-f(x)-f(y)$ does not have compact support in
$E \times E$, and may not be integrable with respect
to the product measure $\nu \otimes \nu$. 

\begin{figure}
\label{fig:unbounded}
\caption{Cancellations in $Tf(x,y)$. Integrals over the shaded regions
diverge absolutely unless $f$ has a suitable modulus of continuity.}
\centerline{\epsfysize=7cm{\epsffile{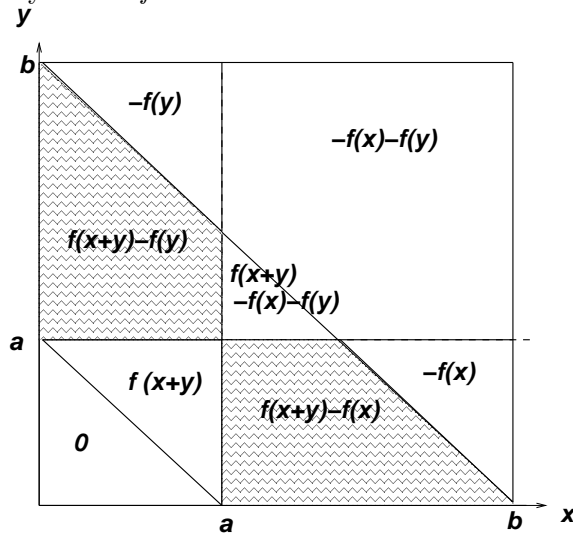}}}
\end{figure}

Here is a counterexample for $K=x+y$.  
Let $\chi(x)$ be the indicator function for
the interval $(0,1)$ and let 
$\nu(dx) = x^{-3/2}\chi(x)dx + \delta(x-2) $. 
Let $f$ be a continuous function with support in
$[a,b]=[2,5]$. Then the values
of $Tf$ are as shown in Figure~\ref{fig:unbounded}. 
Notice that $\nu((x,\infty))$ diverges like $O(x^{-1/2})$ as $x\to0$,
but $\nu$ has finite mass.
Thus,
in order for the integrals in the definition of $L$ to converge, it is
necessary that there be suitably rapid cancellations as we approach
the boundaries. One may show that the integral is finite on all
regions except near the axes in the shaded regions. 
However, we explicitly compute that for any small $\delta>0$,
\begin{eqnarray}
\int_0^\delta\int_0^\infty [f(x+y)-f(x)-f(y)](x+y)\,\nu(dx)\nu(dy)
\qquad\qquad &&\nonumber\\
 =\  \int_0^\delta f(2+y)(2+y)y^{-3/2}dy.
&&\label{eq:counter}
\end{eqnarray}
This is evidently infinite if $f(x)$ rises sufficiently steeply for $x>2$.
Therefore $L(\nu)$ is not a measure.

This means that the space of continuous functions is not appropriate
as a space of test functions in (\ref{eq:weak1}).
The smaller spaces $\mathcal{E}_\gamma$ serve as a suitable
substitute.

\begin{defn}
\label{def:soln_smol}
For each kernel $K(x,y)=1$, $x+y$, $xy$ with degree of
homogeneity $\gamma=0$, 1, 2 respectively,
let $T_\gamma=\infty$ for  $\gamma =0, 1$ and $T_\gamma =\Tgel$ for
$\gamma=2$. 
We say that a map $t \mapsto \nu_t: [0,T_\gamma) \mapsto \mathcal{M}_+$
is a solution to Smoluchowski's coagulation equation if
\begin{enumerate}
\item $\mmt_\gamma(0)=\int_E x^\gamma \nu_0(dx) <\infty$.
\item For all compact sets $B \subset E$, the 
map $t \mapsto \nu_t(B)$ is measurable.
\item $\int_0^t \mmt_\gamma(\tau)^2\,d\tau <\infty$ for all
$t\in(0,T_\gamma)$.
\item For all $f \in {\mathcal{E}_\gamma}$ and $t \in [0,T_\gamma)$ we have
\begin{equation}
\label{eq:weak_soln}
\ip{f, \nu_t} = \ip{f, \nu_0} + \int_0^t \ip{f,L(\nu_\tau)} \,d\tau.
\end{equation}
\end{enumerate}
\end{defn}

\subsection{Existence and uniqueness for the constant kernel}
We will set $K=2$ instead of the usual convention of setting $K=1$,
since it simplifies several calculations (we actually revert to an
older convention, see for example, eq. 459 in \cite{Chandra}).

\begin{thm}
\label{thm:const_well_posed}
Let $\nu_0 \in \mathcal{M}_+$ be a finite measure. 
Then Smoluchowski's coagulation equation with kernel $K=2$
has a unique solution with initial data $\nu_0$, 
and this solution is determined by the solution
of (\ref{eq:evol-const}).
\end{thm}
Theorem~\ref{thm:const_well_posed} is a consequence of~\cite[Thm
2.1]{Norris}. We will prove it anew with the Laplace transform, as 
 the explicit solution formula
is needed later. Let $\num_t$ denote the number distribution at time 
$t$, and $\varphi(t,\cdot)$ be determined from $\num=\num_t$ by 
\qref{eq:renorm_6} for each $t$. Then formally $\varphi$ should
solve the simple equation
(\ref{eq:evol-const}).
For fixed $s >0$, equation (\ref{eq:evol-const}) is an ordinary
differential equation with the solution 
\begin{equation}
\label{eq:const_5}
\varphi(t,s) = \frac{\varphi_0(s)}{1 + \varphi_0(s)t }.
\end{equation}
\begin{lemma}
\label{le:const_well_p}
Assume $\nu_0 \in \mathcal{M}_+$ is finite.  The formula
(\ref{eq:const_5}) determines a weakly continuous 
 map $ [0,\infty) \ni t \mapsto
\nu_t \in \mathcal{M}_+$ with decreasing total number
$\mmt_0(t)=\mmt_0(0)/(1+\mmt_0(0)t)$.
\end{lemma}
\begin{proof}
The solution $\varphi(t,s)$ has  the important property that its
derivative is completely 
monotone for $t \geq 0$. This is because it may be written as a
composition of positive functions with completely monotone derivative
\[ \varphi(t,s) = \frac{p}{1+ tp} \circ (\varphi_0(s)).\]
Recall that the derivative of $\varphi(t,s)$ is $u(t,s)$ the Laplace transform
of the mass measure, say $\mu_t(dx)$. Since $u(t,s)$ is completely
monotone, it follows that $\mu_t(dx) \in \mathcal{M}_+$. Since
$u(t,s)$ is analytic in $t$, we see that the measures $\mu_t$ are 
 weakly continuous  by the classical duality between pointwise
 convergence of the Laplace transform and weak convergence of measures.
That is for any continuous function $f$ with compact support in $E$
we have $\ip{f,\mu_\tau} \rightarrow \ip{f,\mu_t}$ as $\tau \rightarrow
t$. It follows that $\ip{f,\nu_\tau} \rightarrow \ip{f,\nu_t}$ as $\tau
\rightarrow t$ where $\nu_t(dx) = x^{-1} \mu_t(dx)$ is the number
measure.  The statement regarding $\mmt_0(t)$ follows by taking
$s\to\infty$ in (\ref{eq:const_5}).
\end{proof}
\begin{rem}
It is strange at first to consider the desingularized Laplace
transform when the initial data is finite, and indeed the usual
Laplace transform suffices. But (\ref{eq:const_5}) shows us that
the solution is instantly regularizing in the
following sense. If $\num_0(\infty) = \varphi_0(\infty) =\infty$, the
solution satisfies
\[ \lim_{s \to \infty} \varphi(t,s) = \lim_{s \to \infty} 
\frac{\varphi_0(s)}{1+t
  \varphi_0(s)} = \frac{1}{t}.\]
Thus, the number of clusters is finite for $t >0$. Thus,
  $\varphi(t,s)$ defines a natural solution even for an initially 
infinite measure. However,
it is hard to verify (\ref{eq:weak_soln}) in this case (even for
  $f \in C_c^1(E)$), and we
  restrict ourselves to finite measures in what follows. 
\end{rem}
\begin{proof}[Proof of Theorem~\ref{thm:const_well_posed}]
Let us first show that the measures $\nu_t$ determined by the Lemma
are a solution to
Smoluchowski's equation in the sense of Definition~\ref{def:soln_smol}.
Property (1) has been assumed. It is easy to check property (2). The
measures $\nu_t$ are weakly continuous. Thus for a fixed compact set
$B \subset E$, the function $t\mapsto \nu_t(B)$ is semicontinuous.
Property (3) follows from Lemma~\ref{le:const_well_p}.

It is not a priori obvious that $\ip{1-e^{-sx},L(\nu_t)}$ is indeed
$-\varphi(t,s)^2$. But the measures are finite since $\nu_t(E) \leq
\nu_0(E) <\infty$, and thus we may set $f=1-e^{-sx}$ in the definition
of $\ip{f,L(\nu_t)}$ to recover $-\varphi(t,s)^2$.
In particular this shows that 
(\ref{eq:weak_soln}) holds for $f=1-e^{-sx}$.
This equation also holds if $f$ is a monomial $(1-e^{-sx})^k$ because
\[ (1-e^{-sx})^k = \sum_{j=0}^k   \binom{k}{j} (-1)^j e^{-jsx} = 
-\sum_{j=0}^k \binom{k}{j} (-1)^j (1-e^{-jsx}). \]
Given $f \in C(\bar{E})$ and $\varepsilon>0$, Lemma~\ref{le:bernstein}
guarantees an approximation $P_n$ with 
$\|f-P_n\|_{C(\bar{E})} < \varepsilon$. Then,
$ |\ip{f-P_n, L(\nu_t)}| \leq 3 \varepsilon \left(
  \nu_t(E)\right)^2 \leq 3 \varepsilon \left(  \nu_0(E)\right)^2$. Thus, 
\begin{eqnarray}
\nonumber
\lefteqn{\left| \ip{f,\nu_t} - \ip{f,\nu_0} - \int_0^t \ip{f, L(\nu_\tau)}
    \,d\tau \right|}\\
\nonumber
&& \leq \ip{|f-P_n|,\nu_t} + \ip{|f-P_n|,\nu_0} + \int_0^t \left|
  \ip{|f-P_n|,L(\nu_\tau)} \right| \,d\tau \\
\nonumber
&& \leq \varepsilon \left( \nu_0(E) + \nu_0(E) + 3t ( \nu_0(E))^2
\right). 
\end{eqnarray}
This shows that the measures $\nu_t$ define a solution.

Suppose $\nu_t$ and $\tilde{\nu}_t$ are two solutions with the same
initial data. Since $f=1-e^{-sx} \in \mathcal{E}_0=C(\bar{E})$ and
$\varphi(t,s)=\ip{f,\nu_t}\le \mmt_0(t)$ for a.e.\ $t$, we can use
(\ref{eq:weak_soln}) and part (3) of Definition~\ref{def:soln_smol}
to obtain (\ref{eq:evol-add}) in time-integrated form for each
solution.  It follows easily that for fixed $s>0$, each $\varphi(t,s)$ 
is $C^1$ in $t$ and satisfies (\ref{eq:evol-add}).
But this equation has a unique solution $\varphi(t,s)$ as in
(\ref{eq:const_5}). 
As we have noted in Lemma~\ref{le:const_well_p}, 
$\varphi(t,s)$ determines the measure $\nu_t$.
Thus $\nu_t = \tilde{\nu}_t$.
\end{proof}

\subsection{Existence and uniqueness for the additive kernel}
We always work with solutions of finite mass, but
we do not assume that the
number of clusters is finite. Therefore, 
$L(\nu)$ will not be a measure. Nevertheless, it does define a bounded linear
functional on the space of Lipschitz functions on $E$. 
\begin{lemma}
\label{le:lipschitz}
Let $\nu \in \mathcal{M}_+$ with $\mmt_1 := \int_E x\,\nu(dx) <
\infty$, and let $L(\nu)$ be defined by (\ref{eq:defn_L}) with
$K=x+y$. Suppose $f$ is Lipschitz and $f(0)=0$. Then
\begin{equation}
\label{eq:defn_L2}
|\ip{f,L(\nu)}| \leq 2 \mmt_1^2 \Lip(f). 
\end{equation}
\end{lemma}
\begin{proof}
By the symmetry of the integral in (\ref{eq:defn_L}), we see that
\[ |\ip{f,L(\nu)}| \leq \int_E \int_E |f(x+y)-f(x)-f(y)| y
\,\nu(dy) \nu(dx).\]
The integrand is controlled by
\[ |f(x+y)-f(x)-f(y)| \leq |f(x+y)-f(y)| + |f(x)| \leq 2\Lip(f) x.\]
Thus we obtain
\[ |\ip{f,L(\nu)}| \leq 2\Lip(f) \int_E \int_E x y \,\nu(dy) \nu(dx) =
2\mmt_1^2 \Lip(f).\]
\end{proof}

\begin{thm}
\label{thm:add_well_posed}
Let $\nu_0 \in \mathcal{M}_+$ satisfy $\int_E x
\,\nu_0(dx)= \mmt_1<\infty$. Then Smoluchowski's 
coagulation equation with kernel $K=x+y$
has a unique solution with initial data $\nu_0$, such that $\int_E x
\,\nu_t(dx) =\mmt_1$ for all $t \in [0,\infty)$.
\end{thm}
We will construct a solution using the desingularized 
Laplace transform and then
prove its uniqueness. The evolution equation for $\varphi$ is 
\[ \D_t \varphi -\varphi \D_s\varphi = -\mmt_1 \varphi.\]
We may always normalize initial data such that
$\mmt_1=1$, and we assume this in all that follows. Thus, we have
\begin{equation} 
\label{eq:evol-add2}
\D_t \varphi -\varphi \D_s\varphi = -\varphi
\end{equation}
It is striking that (\ref{eq:evol-add2}) is
simply the inviscid  
Burgers' equation with linear damping. However, there is no shock
formation, since the initial data are analytic with a completely
monotone derivative 
satisfying $\D_s\varphi_0(s) \leq 1$. This can be seen in the explicit
solution below which is valid for all time.  Since $u = \partial_s
\varphi$, differentiating (\ref{eq:evol-add2}) we have
\begin{equation}
\label{eq:evol-add-u}
\partial_t u - \varphi  \partial_s u = -u(1-u). 
\end{equation}
We solve (\ref{eq:evol-add2}) and (\ref{eq:evol-add-u})
globally by the method of characteristics. Let $s(t,\sigma)$ denote
the characteristic that originates at 
$\sigma$ at $t=0$\/. Then we have 
\begin{equation}
\label{eq:add4}
\frac{ds}{dt} = - \varphi , \quad
\frac{d\varphi}{dt} = -\varphi, \quad \frac{du}{dt} = -u(1-u).
\end{equation}
on a characteristic. We integrate (\ref{eq:add4}) along the
characteristics  to obtain 
\begin{eqnarray}
\label{eq:add_soln1}
\varphi(t,s) & =&  e^{-t} \varphi_0(\sigma), \\
\label{eq:add_soln2}
s(t,\sigma) - \varphi(t,s) & = &\sigma - \varphi_0(\sigma),\\
\label{eq:add_soln3}
u(t,s) &= &\frac{e^{-t}u_0(\sigma)}{1-(1-e^{-t})u_0(\sigma)}.
\end{eqnarray}
\begin{lemma}
\label{le:add_well_p}
Suppose $\nu_0(x) \in \mathcal{M}_+$ with $\int_E x \,\nu_0(dx)=1$. 
Then equation (\ref{eq:evol-add2}) determines a
map  $ [0,\infty) \ni t \mapsto
\nu_t \in \mathcal{M}_+$  such that 
\begin{enumerate}
\item $\int_E x \nu_t (dx)=1$ for all $t$.
\item $\mu_t = x\nu_t$ is weakly continuous.
\end{enumerate}
\end{lemma}
\begin{proof}
Observe that
\[ \sigma - \varphi_0(\sigma) = \int_E(e^{-sx} - 1 +sx)
\,\num_0(dx) > 0, \quad \sigma >0. \]
Thus, by (\ref{eq:add_soln1}) and (\ref{eq:add_soln2}) 
\begin{equation}
\label{eq:CM1}
 s(t,\sigma) = \sigma - \varphi_0(\sigma)(1-e^{-t}) >0, \quad \sigma
>0.
\end{equation}
The right hand side is a strictly increasing  function of $\sigma$ for
all $t \geq 0$, thus, the inverse map $\sigma(t,s)$ is well
defined. Differentiating (\ref{eq:CM1}) with respect to $s$ we find
that
\begin{equation}
\label{eq:CM2}
\frac{d\sigma}{ds} = \frac{1}{1-(1-e^{-t})u_0(\sigma(s))}, \quad \mbox{whence} \quad u(t,s) = e^{-t} u_0(\sigma)\frac{d\sigma}{ds}.
\end{equation}
Since $u_0$ is the Laplace transform of a positive measure, it is a
completely monotone function of $\sigma$. In order to show that
$u(t,s)$ is completely monotone in $s$, it suffices to show that
$d\sigma/ds$ is completely  monotone in $s$ (see Criterion
1 and 2 in~\cite[XIII.4]{Feller}). We prove this as follows. Let us 
consider the sequence of iterates 
\[ \sigma_0(s)=s \quad \mbox{and} \quad \sigma_{n+1}(s) = s +
(1-e^{-t}) \varphi_0(\sigma_n(s)), \;\;n \geq 0.\] 
Clearly, $|\sigma_{n+2}(s)-\sigma_{n+1}(s)| < |\sigma_{n+1}(s)-
\sigma_n(s)|$ so that $\sigma_{n}(s) \rightarrow \sigma(s)$ the unique
solution to (\ref{eq:CM1}). Moreover,  we have   
\[ \frac{d\sigma_{n+1}(s)}{ds} = 1 + (1-e^{-t}) u_0(\sigma_n(s))
\frac{d\sigma_n(s)}{ds}. \] 
Thus if  $d\sigma_{n}/ds$ is completely monotone, then so is
$d\sigma_{n+1}/ds$. But $d\sigma_0/ds =1$ is completely monotone. By
induction, 
$d\sigma_{n}/ds$ is completely monotone for $n \geq 1$ and so is the
limit $d\sigma/ds$.  

We may now conclude that 
the solution $\varphi(t,s)$ to
(\ref{eq:evol-add2}) defined by (\ref{eq:add_soln1}) exists for all $t
\geq 0$, is unique, and has a completely monotone derivative
$u(t,s)$. $u(t,s)$ defines a unique mass measure, say
$\mu_t(dx)$. We see from the solution (\ref{eq:add_soln3})
 that $u(t,0)=u_0(0)=1$. Thus, 
the total mass $\int_E \mu_t(dx)= u(t,0)= 1$
for all $t \geq 0$. The measures $\mu_t$ are 
weakly continuous since $u(t,s)$ is analytic in time. 
\end{proof}
\begin{proof}[Proof of Theorem~\ref{thm:add_well_posed}]
Let us first check that the measures $\nu_t$ determined by
(\ref{eq:evol-add2}) solve Smoluchowski's
equation in the sense of Definition~\ref{def:soln_smol}.
Conditions (1) and (2) in Definition~\ref{def:soln_smol} are verified
as in the proof of Theorem~\ref{thm:const_well_posed}. Since 
$\nu_t$ has constant mass, it follows from Lemma~\ref{le:lipschitz}
that the functionals $L(\nu_t)$ are uniformly bounded on $\mathcal{E}_1$. In
particular 
\begin{equation}\label{eq:L-add}
\ip{1-e^{-sx}, L(\nu_t)} = -\varphi + \varphi \D_s\varphi
\end{equation}
as desired. This shows that (\ref{eq:weak_soln}) holds for
$f=1-e^{-sx}$, and thus for monomials $(1-e^{-sx})^k$.
Given any $f \in \mathcal{E}_1$ and $\varepsilon >0$ we
choose an approximation $P_n(x;s)$ as in Lemma~\ref{le:bernstein}(b) so
that $\|f-P_n\|_{C^1(\bar{E})} < \varepsilon$. Then
by Lemma~\ref{le:lipschitz}
\[ |\ip{f-P_n, L(\nu_t)}| \leq 2\Lip(f-P_n) < 2 \varepsilon.\]
Similarly, 
\[ |\ip{f-P_n,\nu_t}| \leq \int_E |f(x)-P_n(x;s)| \,\nu_t(dx) \leq \Lip(f-P_n)
\int_E x \,\nu_t(dx)  <  \varepsilon. \]
This shows that $\nu_t$ is a solution in the sense of
Definition~\ref{def:soln_smol}. 

Now suppose only that ${\nu}_t$ is a solution in the sense of 
Definition~\ref{def:soln_smol}. 
The function $f = 1-e^{-sx} \in \mathcal{E}_1$, 
and $f\le sx$. For a.e.\ $t>0$ we have 
$\mmt_1(t)<\infty$ and from 
$\varphi=\ip{f,{\nu}_\tau}\le s^{-1} \mmt_1(t)$
it follows $\varphi(t,s)$ is analytic in $s$ with
$|\D_s^k\varphi|\le s^{1-k}\mmt_1(t)$ for $k=1,2,\ldots$.
Part (3) of Definition~\ref{def:soln_smol} 
ensures that for fixed
$s>0$, $\mmt_1(t)$ 
is locally square-integrable on $[0,\infty)$ and 
(\ref{eq:weak_soln}) gives (\ref{eq:evol-add2}) in time-integrated
form, since (\ref{eq:L-add}) holds for a.e.\ $t$ and 
\[
|\ip{f,L(\nu_t)}|=|\varphi\D_s\varphi-\varphi|\le
s\mmt_1(t)(\mmt_1(t)+1).
\]
It follows that $\varphi$ is 
continuous in $t$, uniformly for $s$ in compact sets in $E$.
Moreover, one can justify differentiating (\ref{eq:weak_soln}) in $s$ 
and infer that $\D_s\varphi$ is continuous.
Then $\varphi$ is a $C^1$ solution of (\ref{eq:evol-add2}) whence 
$\varphi(t,s)$ is uniquely determined by initial data. But as
we have noted in Lemma~\ref{le:add_well_p}, $\varphi(t,s)$ uniquely
determines the measure $\nu_t$. Thus the solution is unique.
\end{proof}

\subsection{Existence and uniqueness for the multiplicative kernel}
\label{subsec:mult}
The multiplicative kernel differs from the constant and additive
kernels, since it is not well posed for all time. But, the analysis 
can be formally reduced to the additive case by
a change of variables. This is
well-known~\cite{Drake}, but we include it for completeness.

The divergence of the classical self-similar solution is $O(x^{-5/2})$
as $x \rightarrow 0$. The total number and mass are infinite, but
the second moment is finite. Therefore, we consider the following
desingularized Laplace transform
\begin{equation}
\label{eq:defn_phi}
\phi(t,s) = \int_E (e^{-sx}-1+sx) \,\nu_t(dx).
\end{equation} 
We substitute $f=e^{-sx}-1+sx$ in the equation of evolution
(\ref{eq:defn_L}) to find
\begin{equation}
\label{eq:evol-phi}
 \partial_t \phi = \ip{f, L(\nu_t)} = \frac{1}{2} ( \partial_s  \phi)^2. 
\end{equation}
Equation (\ref{eq:evol-phi}) is the Hamilton-Jacobi equation
associated to the inviscid Burgers equation. Thus, we let
\begin{equation}
\label{eq:defn_psi}
\psi(t,s) = \partial_s \phi = \int_E (1-e^{-sx}) x \,\nu_t(dx).
\end{equation}
Then from (\ref{eq:evol-phi}) we have 
\begin{equation}
\label{eq:mult3b}
\partial_t \psi -\psi \partial_s \psi =0.
\end{equation}
The exact solution to (\ref{eq:mult3b}) 
with initial data $\psi_0(s)$ may be found by the
method of characteristics. The characteristic originating at $s_0$ is
denoted 
\[ s(t,s_0) = s_0 - \psi_0(s_0) t. \]
Let $t(s_0,s_1)$ denote the time for two characteristics originating at $s_0 <
s_1$ to intersect. Then, if this is the first intersection
\[ \frac{1}{t}= \frac{\psi_0(s_1) -\psi_0(s_0)}{s_1 -s_0},
\quad{\mbox{whence}}\quad \frac{1}{\partial_s \psi_0(s)} <
t(s_0,s_1) < \frac{1}{\partial_s \psi_0(s_1)}, \]
where the inequalities follows from the mean value theorem and the
complete monotonicity of $\partial_s \psi_0$. Thus, letting $s_0=0$
and $s_1 \rightarrow 0$, we see that the least time taken for
characteristics to intersect is given by
\[ \Tgel^{-1} = \partial_s \psi_0 (0) = \int_E x^2 \,\nu_0(dx)
=\mmt_2(0).\]
Without loss of generality, we may assume that
the initial data is normalized so that $\mmt_2(0)=1 =
\Tgel$ and thus $0 \leq t <1$. 
This normalization assumption is analogous to the assumption
that $\mmt_1=1$ for the additive kernel. 
Equation (\ref{eq:mult3b})
should be compared with  equation (\ref{eq:evol-add}). In 
fact, given initial data $\psi_0(s)$,
by changing the time scale in (\ref{eq:evol-add}) it is easy to
check that $\psi(t,s)$ is the solution to (\ref{eq:mult3b}) if and only
if 
\begin{equation}
\label{eq:mult4}
\psi(t,s) = \frac{1}{1-t} \varphi(-\log(1-t),s),
\end{equation}
where $\varphi(t,s) $ is the unique solution to (\ref{eq:evol-add}) with
initial data $\psi_0$. The next lemma follows
immediately from Lemma~\ref{le:add_well_p}.
\begin{lemma}
\label{le:mult_well_p}
Suppose $\num_0(x) \in \mathcal{M}_+$ with $\int_E x^2 \,\num_0(dx)=1$ 
Then equation (\ref{eq:mult3b}) determines a map  $ [0,1) \ni t \mapsto
\nu_t \in \mathcal{M}_+$ such that
\begin{enumerate}
\item $\mmt_2(t)= \int_E x^2 \,\nu_t(dx) = (1-t)^{-1}$.
\item $x^2 \nu_t$ is weakly continuous on $[0,1)$.
\end{enumerate}
\end{lemma}
It is natural to term the measure $\nu_t$ the solution to
Smoluchowski's coagulation equation with kernel $K=xy$. However, it is
harder to formulate a completely natural well-posedness theory in this case,
and we will settle for a reasonable compromise.
\begin{defn}
Define the norm 
\begin{equation}
\label{eq:defn_V} \sup_{x,y>0} \frac{|f(x+y)-f(x)-f(y)|}{xy} :=
\|f\|_V, 
\end{equation}
and the associated Banach space $V = \{f \in C_0(E)| \|f\|_V < \infty \}$. 
\end{defn}
It is clear that $V$ is a Banach space.  The norm $\|\cdot\|_V$ is
natural in the following sense. 
\begin{lemma}
\label{le:lipschitz2}
Let $\nu \in \mathcal{M}_+$ such that $\mmt_2<\infty$, and let $L(\nu)$
be defined by (\ref{eq:defn_L}) with $K=xy$.
Then $L(\nu)$ defines a bounded linear functional on $V$ with
norm $\leq \mmt_2^2/2$.
\end{lemma}
\begin{proof}
Since $|f(x+y)-f(x)-f(y)| \leq \|f\|_V xy$ we have
\[ |\ip{f,L(\nu)}| \leq \frac{1}{2} \int_E \int_E \|f\|_Vx^2 y^2 \,\nu(dx)
\nu(dy) = \frac{\mmt_2^2}{2} \|f\|_V.\]
\end{proof}
It is easy to check that finite sums
$f(x) = \sum_{k=1}^n a_k (1-e^{-s_k x})$ are in $V$. 
We would like to believe that these functions are dense in $V$,
but this seems hard to prove, as the norm above is 
unwieldy. Instead we will work with $C^2$ functions and use
the following, whose easy proof we omit.
\begin{lemma}
\label{le:C2_V}
Let $f$ be a $C^1$ function such that $f(0)=0$ and $f'$ is 
Lipschitz. Then $\|f\|_V \leq 2\Lip(f')$.
\end{lemma}

\begin{thm}
\label{thm:mult_well_posed}
Let $\nu_0 \in \mathcal{M}_+$ satisfy $\mmt_2(0)<\infty$. Then Smoluchowski's 
coagulation equation with kernel $K=xy$
has a unique solution with initial data $\nu_0$ on the time interval
$[0,\mmt_2(0)^{-1})$.
\end{thm}
\begin{proof}
Without loss of generality we may suppose that $\mmt_2(0)=1$.
The measures $\nu_t$ of Lemma~\ref{le:mult_well_p} 
are a candidate solution, and it is easy to check
that conditions (1) and (2) of Definition~\ref{def:soln_smol} are satisfied. 
Since $\mmt_2(t) = (1-t)^{-1}$ by Lemma~\ref{le:lipschitz2} 
one sees that $L(\nu_t)$ is a bounded linear operator on $V$ and
in particular
\[ \ip{1-sx-e^{-sx},L(\nu_t)} = -\frac{1}{2} \left( \phi_s
\right)^2.\]
Thus, $\phi(t,s)$ solves~(\ref{eq:evol-phi}). Yet, some care is
needed in checking that (\ref{eq:weak_soln}) holds in full generality. 
Let $f \in \mathcal{E}_2$. 
We apply Lemma~\ref{le:bernstein} to $f'$ (notice, {\em not\/}
$f$) to obtain an approximation $P_n(x;s)$ with 
$\sup_x |P_n-f'| < \varepsilon$, and $\sup_x
|P'_n-f''| < \varepsilon$. Observe that, we may rewrite
\[ P_n(x;s) = \sum_{k=1}^n a_{n,k} (1-e^{-sx})^k = \sum_{k=1}^n
b_{n,k}(1-e^{-skx}) \]
by expanding $(1-e^{-sx})^k$ with the binomial formula, and defining
$b_{n,k}$ as the corresponding linear combinations of $a_{n,k}$. We integrate
$P_n$ to obtain the approximation
\[ Q_n(x;s) = \sum_{k=1}^n \frac{b_{n,k}}{sk} \left(e^{-skx}-1+skx
\right).\]
Notice that $P_n(0;s)=0$. Therefore, by the fundamental theorem of
calculus, we also have
\[
 |P_n(x;s)-f'(x)| \leq \int_0^x |P'_n(z;s)-f''(z)|dz < \varepsilon
x,
\]
and upon integration again,
\[ |f(x) - Q_n(x;s)| < \frac{\varepsilon x^2}{2}. \]
But we then have,
\[ \left| \ip{f-Q_n,\nu_\tau} \right| \leq \frac{\varepsilon}{2}
\int_E x^2 \nu_\tau(dx) = \frac{\varepsilon}{2}\mmt_2(\tau), \quad \tau
\in [0,1).\] 
Since $\sup_x|f''-Q_n''| < \varepsilon$ we apply
Lemma~\ref{le:lipschitz2} and  Lemma~\ref{le:C2_V} to obtain
\[ \left| \ip{f-Q_n, L(\nu_\tau)} \right| < 
\varepsilon \mmt_2^2(\tau), \quad \tau \in
[0,1).\]
This proves that $\nu_t$ is a solution.

It is slightly harder to prove uniqueness in this case. 
As in Theorem~\ref{thm:const_well_posed} and
Theorem~\ref{thm:add_well_posed} it suffices to deduce uniqueness of
the measure valued solution via uniqueness of solutions to
(\ref{eq:evol-phi}). The obtstruction is that it is not clear from the
definition of the weak solution that (\ref{eq:evol-phi}) holds, 
since the test functions $1-sx-e^{-sx}$ do not lie in
$\mathcal{E}_2$. This can be overcome with an approximation argument
that we only sketch. We consider  $f_n(x)= (1-sx-
e^{-sx}) \chi_n(x)$ where $\chi_n$ is a $C^\infty$ cut-off 
function such that $\chi_n =1, x \leq n, \chi_n =0, x \geq n+1$.
By the monotone convergence theorem, $\lim_{n \to \infty} \ip{f_n,\nu_t}
= \ip{1-sx-e^{-sx}, \nu_t}$. By Lemma~\ref{le:lipschitz2} and
Lemma~\ref{le:C2_V}
$\ip{f_n,L(\nu_t)}$ is well-defined and uniformly
bounded by $C_s\mmt_2(t)^2$. We may then use
the dominated convergence theorem to deduce that (\ref{eq:evol-phi})
holds in the limit $n \to \infty$. Uniqueness of $\nu_t$ follows.
\end{proof}

\section{Regular variation}
\label{sec:rv}
Several formal calculations by physicists working on Smoluchowski's
equations take the following form : (1) assume that the number density
$n(x) \sim x^{\alpha}$ for some scaling exponent $\alpha$, and (2) 
conclude based on physical arguments that $\alpha$ takes a particular value.  
The theory of regular variation helps us makes these formal
calculations precise, and lays bare the mechanism controlling
the approach to scaling form.  Our primary source is  Feller's book, and
we restate below useful results from~\cite[VIII.8]{Feller}. The theory
of regular variation has many applications in analysis and probability, and an
authoritative text, rich in examples, is~\cite{Bingham}.

\subsection{Rigidity of scaling limits}
Loosely speaking, a function is {\em slowly varying\/} if it is 
asymptotically flat under changes of scale. Precisely, we say that
a positive function $L(x)$ is {\em slowly varying at infinity\/} if 
\begin{equation}
\label{eq:rv1}
\lim_{x \to \infty} \frac{L(tx)}{L(x)} = 1, \quad\mbox{for all $t >0$}.
\end{equation}
For example, all powers and iterates of $\log x$ are slowly varying at
infinity. If we consider the limit $x \to 0$ instead, we obtain functions
that are slowly varying at $0$.   

A function $N(x)$ is {\em  regularly varying at infinity\/}  with index 
$\rho \in \mathbb{R}$ if there is a slowly varying function $L(x)$ such that
\begin{equation}
\label{eq:rv2}
N(x) \sim x^\rho
L(x)\;\;\mathrm{as}\;\;{x \to \infty} . 
\end{equation}
The notation $\sim$ means $\lim_{x \to \infty} N(x)/x^\rho L(x) =1$. 

The notion of regular variation is intimately related to necessary
and sufficient condtions for the existence of scaling limits.
This is reflected in the following classical ``rigidity'' 
lemma~\cite[Lemma VIII.8.3]{Feller}, which will be one of our principal tools.

\begin{lemma}
\label{le:rv}
Suppose that 
${a_{n+1}}/{a_n} \to 1$ {and} $\lambda_n \to \infty$ as $n\to\infty$.
If $\varphi$ is a positive, monotone function such that 
\[ \lim_{n \to \infty} a_n \varphi\left(\frac{s}{\lambda_n}\right) = g(s) \leq
\infty \] 
exists for $s$ in a dense subset of $(0,\infty)$, and $g$ is finite
and positive on some 
interval, then $\varphi$ varies regularly at $0$ and $g(s) = cs^\rho$
for $-\infty < \rho < \infty$ and some $c >0$. 
\end{lemma}

\subsection{Tauberian theorems}
We will rigorously deduce the
asymptotics of $\num$ by the beautiful
Hardy-Littlewood-Karamata Tauberian theorem~\cite[XIII.5]{Feller}. 
\begin{thm}
\label{thm:tauber}
If $L$ is slowly varying at infinity and $0 \leq \alpha < \infty$, then the
following are equivalent:
\[ \num(x) \sim {x^\alpha L(x)} \quad \mbox{as $x \to \infty$} ,\] 
and
\[ \eta(s) \sim s^{-\alpha} L\left(\frac{1}{s}\right)\Gamma(1+\alpha)  \quad
\mbox{as $s \to 0$}.\] 
Moreover, this equivalence remains true when we interchange the roles of the
origin and infinity, namely when $s \to \infty$ and $x \to 0$.
\end{thm}
We will use the following lemma to show that there is no loss of
generality in working with $\varphi$ instead of $\eta$.
\begin{lemma}
\label{le:diff_phi}
Suppose $\partial_s \psi$ is the Laplace transform of a positive
measure. Let $\alpha <1$ and $L$ be a function slowly varying at $0$.
The following are equivalent.
\begin{enumerate}
\item $\psi(s)-\psi (0) \sim s^{1-\alpha} L(s)$ as $s \to 0$.
\item $\partial_s \psi(s) \sim (1-\alpha) s^{-\alpha} L(s)$ as $s \to 0$.
\end{enumerate}
\end{lemma}
\begin{proof}
Suppose $(1)$. Since $\psi(s)-\psi(0) = s^\alpha L(s) h(s)$ with $\lim_{s
  \to 0} h(s) =1$, without loss of generality we may write $\psi(s)-\psi(0)
  = s^\alpha L(s)$.  Fix $a >1$. Then by the mean value theorem and
  the complete monotonicity of $\partial_s \psi$ we have
\[s(a-1) \partial_s \psi (s) \geq \psi(as) - \psi(s) = s^{1-\alpha}
  L(s) \left( a^{1-\alpha} \frac{L(as)}{L(s)} -1 \right). \]
Thus, letting $s \to 0$, and using (\ref{eq:rv1})  we have
\[ \liminf_{s \to 0} \frac{s^\alpha \partial_s \psi (s)}{L(s)}  \geq
\frac{a^{1-\alpha}-1}{a-1}. \]
Since $a >1$ is arbitrary, we may maximize the right hand side to
obtain
\[ \liminf_{s \to 0} \frac{s^\alpha \partial_s \psi (s)}{L(s)} \geq
1-\alpha.\]
Choosing $a <1$ and using a similar argument yields,
\[ \limsup_{s \to 0} \frac{s^\alpha \partial_s \psi (s)}{L(s)} \leq
1-\alpha.\]
Thus, $\partial_s \psi (s) \sim (1-\alpha) s^{-\alpha} L(s)$. 

Conversely, assume (2). Then we have
\[ \psi(s)-\psi(0) = (1-\alpha) \int_0^s t^{-\alpha}L(t) dt
=(1-\alpha) s^{1-\alpha} L(s) \int_0^1 t^{-\alpha}\frac{L(st)}{L(s)}\,dt.
\]
Since $L$ is slowly varying at zero, then for any constants
$A>1$ and $\delta>0$ there exists $s_0$ such that for $0<s\le s_0$,
$0<t\le1$ we have $L(st)/L(s) \le A t^{-\delta}$ 
(this is not hard to show, but see Theorem 1.5.6 in \cite{Bingham}). 
Then (1) follows by dominated convergence.
\end{proof}

\section{Scaling solutions for the constant kernel}

\subsection{Mittag-Leffler distributions}
The scaling solution 
\begin{equation}
\label{eq:scaling1}
n(t,x) = t^{-2}\exp ({-x}/{t}),\quad t >0 
\end{equation}
is the continuous limit of a special solution found by 
Smoluchowski~\cite{Aldous}. 
Kreer and Penrose proved that the rescaled number density $t^2n(t,xt)$
converges uniformly to $e^{-\alpha x}$ on compact sets, under 
the assumption that the initial number density $n_0(x)$  be $C^2$ and
have exponential 
decay in $x$~\cite{KP}. The constant $\alpha$ is determined by the initial mass.

In this section we show that the solution (\ref{eq:scaling1}) is just
one of a one-parameter family of scaling
solutions given by 
\begin{equation}
\label{eq:scaling2}
n(t,x) = t^{-1-1/\rho} n_\rho( x t^{-1/\rho}), \quad
t>0, \;\rho \in (0,1], 
\end{equation}
where $n_\rho(x) = \fzero'(x)$ is the density, and $\fzero$ the
distribution function for  the Mittag-Leffler
distribution
\begin{equation}
\label{eq:mittag_leffler}
 \fzero(x) = \sum_{k=1}^\infty \frac{(-1)^{k+1} x^{\rho
 k}}{\Gamma(1+\rho k)} 
 , \quad \rho \in (0,1]. 
\end{equation}
Of these solutions, only the solution (\ref{eq:scaling1}) for $\rho=1$ has
finite mass, and the others have fat tails. 
The Mittag-Leffler distribution was
studied by Pillai~\cite{Pillai} who showed that 
these distributions are infinitely divisible and geometrically
infinitely divisible for $\rho \in (0,1]$. 

For our purposes, it is especially relevant that
the Mittag-Leffler distribution has Laplace transform
\begin{equation}
\label{eq:ML_trans}
 \int_0^\infty e^{-sx} n_\rho(x)dx  = \int_0^\infty e^{-sx} \fzero(dx)
= \frac{1}{1+s^\rho}. 
\end{equation}
In terms of the Mittag-Leffler function
\[
 E_\rho(x)= \sum_{k=0}^\infty \frac{x^k}{\Gamma(1+\rho k)}
\]
one has $F_\rho(x)=1-E_\rho(-x^\rho)$. 
It is interesting and useful to note the following 
recent calculation of Tsoukatos~\cite{Tsoukatos}.
\begin{lemma}\label{Erho}
For $0<\rho<1$ we have
\begin{equation}\label{eq:Erho}
E_\rho(-x^\rho) = \frac1\pi \int_0^\infty e^{-rx} 
\frac{r^{\rho-1}\sin\pi\rho}{(r^\rho+\cos\pi\rho)^2+(\sin\pi\rho)^2}
\,dr.\end{equation}
Hence $E_\rho(-x^\rho)$ and $n_\rho(x)=-\D_xE_\rho(-x^\rho)$
are completely monotone.
\end{lemma}
\begin{proof}
We sketch the argument of Tsoukatos~\cite{Tsoukatos}.
Since (\ref{eq:ML_trans}) implies
\begin{equation}
\int_0^\infty e^{-sx}E_\rho(-x^\rho)\,dx =
\frac{1}{s}\left(1-\frac{1}{1+s^\rho}\right),
\end{equation}
one can invoke the Laplace inversion formula and evaluate
it by deforming the contour to fold along the negative real axis 
to obtain, for any $\sigma>0$,
\[
E_\rho(-x^\rho) = \frac1{2\pi i}\int_{\sigma-i\infty}^{\sigma+i\infty}
e^{sx} \frac{s^\rho}{1+s^\rho}\frac{ds}s 
= \frac1\pi \Im \int_0^\infty e^{-rx}\frac
{r^\rho e^{i\pi\rho}}
{1+r^\rho e^{i\pi\rho}} \frac{dr}r,
\]
and the result follows.
\end{proof}
Note that the complete monotonicity of $E_\rho(-x)$ 
(conjectured by Feller and proved by Pollard in 1948)
also implies the complete monotonicity above, 
since $x^\rho$ is positive with completely monotone derivative.
A point of confusion in the literature is that the term 
``Mittag-Leffler law'' is used by some for the distribution whose
Laplace transform is $E_\rho(-s)$~\cite{Bingham}.

\subsection{Scaling solutions}
\label{sec:const_scaling}
Let us now check that $n_\rho(t,x)$ defined by (\ref{eq:scaling2}) are indeed
solutions to (\ref{eq:smol1}) when $K=2$. We take the
Laplace transform of (\ref{eq:smol1}), and its limit at $s=0$, to obtain
\begin{equation}
\label{eq:const_1}
\frac{\partial \eta}{\partial t} = \eta^2 - 2\eta(t,0) \eta, \quad
\frac{\partial \eta(t,0)}{\partial t} = -\eta(t,0)^2. 
\end{equation}
In analogy with (\ref{eq:scaling1}) we make the ansatz
$ \eta(t,s) = t^{-1} \eta_\rho(s \lambda(t))$, $\eta_\rho(0)=1$
in equation (\ref{eq:const_1}). Letting  $\xi = s \lambda$ we have
\begin{equation}
\label{eq:const_ansatz2}
\left( t \frac{\dot \lambda}{\lambda} \right) \xi \eta_\rho' 
= - \eta_\rho(1 - \eta_\rho). 
\end{equation}
Equation(\ref{eq:const_ansatz2}) may be simplified by separating
variables. We let $\rho$ denote the separation constant to obtain
\begin{equation}
\label{eq:const_ansatz3}
\xi \eta_\rho'  =  -\rho \eta_\rho(1- \eta_\rho), \quad
\frac{\dot \lambda}{\lambda}  =  \frac{1}{\rho t} 
\end{equation}
The general solution to (\ref{eq:const_ansatz3}) is 
\begin{equation}
\label{eq:const_ansatz4}
\eta_\rho(\xi) = \frac{1}{1 + c_1 \xi^\rho}, \quad
\lambda(t) =  (c_2 t)^{1/\rho} 
\end{equation}
where $c_1$, $c_2 >0$ are arbitrary constants. We combine  the two solutions 
to find for each $\rho$, a family of solutions related by scaling in
$s$ and $t$, 
\begin{equation}
\eta(t,s) = t^{-1} \frac{1}{ 1 + c_1 c_2 s^\rho t}, \quad t >0.
\end{equation}
$\eta(t,s)$ is completely monotone if and only if $\rho \in
(0,1]$~\cite{Pillai}, 
thus it is only for $\rho \in (0,1]$ that we obtain positive solutions
to equation (\ref{eq:smol1}). By a trivial scaling we may achieve
$c_1=c_2=1$, and then $n(t,x)$ is given by
(\ref{eq:scaling2}). The scaling solutions have finite mass only when
$\rho=1$, and in this case the mass is conserved.

\subsection{Asymptotics of scaling solutions}
We may use equation (\ref{eq:mittag_leffler}) to
obtain the convergent expansion
\begin{equation}
\label{eq:mittag_leffler2}
n_\rho(x) =  \sum_{k=1}^\infty \frac{(-1)^{k+1} 
  x^{\rho k-1}}{\Gamma(\rho k)},\quad x>0
\end{equation}
which implies the divergence $n_\rho(x) \sim x^{\rho
  -1}/\Gamma(\rho)$ as $x \rightarrow 0+$, for $\rho \in (0,1)$.

Since $n_\rho(x)$ is completely monotone for $\rho\in(0,1)$, 
its asymptotic properties as $x \rightarrow \infty$ may
be obtained rigorously by differentiating the formula in
Lemma~\ref{Erho} and using the Tauberian theorem. 
We obtain
\begin{equation}
\label{eq:const_tauber_2}
n_\rho(x) \sim 
{x^{-\rho-1}}\Gamma(2+\rho)\frac{\sin\pi\rho}{\pi(1+\rho)}
= \frac{ x^{-\rho-1}}{-\Gamma(-\rho)},
\quad\text{as $ x \rightarrow \infty$,}
 \end{equation}
using $y=1+\rho$ in the identity 
\begin{equation}\label{eq:Gamma1}
\Gamma(1+y)\Gamma(1-y) \frac{\sin\pi y}{\pi y} = 1.
\end{equation}
From (\ref{eq:const_tauber_2}), or because
$-\eta_\rho'(s) \sim \rho s^{\rho-1}$ as $s \rightarrow 0$, we find
\begin{equation}
\label{eq:const_tauber_1}
\int_0^x y n_\rho(y) dy \sim \frac{\rho x^{1-\rho}}{\Gamma(2-\rho)},
\quad\text{as $ x \rightarrow \infty$.}
 \end{equation}
The case $\rho =1/2$ curiously admits the exact solution
(see~\cite[Ch. 29]{Abramowitz})
\[ n_\rho(x) = \frac{1}{\sqrt{\pi x}} - e^x \mathrm{erfc} \sqrt{x}.\]

\section{Weak convergence for the constant kernel}
Let $\nu_t$ be the measure-valued solution of Smoluchowski's equation
with kernel $K=2$ obtained in section 2.2, given
initial size distribution $\nu_0$ that is a finite measure.
We normalize to define a {\em probability} distribution
function (the size-biased distribution)
\begin{equation}
\label{eq:rescale_nu}
F(t,x) = \frac{\num_t((0,x])}{\num_t((0,\infty))}, \quad{t >0}.
\end{equation}
We then have the following characterization of
permissible limits  under rescaling and their domains of attraction. 
Below, we call a probability distribution function $F_*(x)$
{\it nontrivial} if $F_*(x)<1$ for some $x>0$, meaning the
distribution is proper ($\lim_{x\to\infty}F(x)=1$) 
and not concentrated at zero.
\begin{thm}
\label{thm:const}
\begin{enumerate}
\item Suppose there is a rescaling function $\lambda(t) \to \infty$ and a 
nontrivial probability distribution function $F_*(x)$ such that
\begin{equation}
\label{eq:hyp1}
\lim_{t\to\infty} F(t,\lambda(t) x) = F_*(x)
\end{equation}
at all points of continuity of $F_*$. Then, there exists $\rho \in (0,1]$ 
and a function $L$ slowly varying at infinity such that
\begin{equation}
\label{eq:hyp2}
\int_0^x y \num_0(dy) \sim {x^{1-\rho} L(x)}
\quad\text{as $x \to \infty$.} 
\end{equation}
\item  Conversely, suppose there exists $\rho \in (0,1]$ and a
function $L$ slowly varying at infinity such that (\ref{eq:hyp2})
holds. Then it follows that there 
is a strictly increasing rescaling $\lambda(t) \to \infty$ such that
\[\lim_{t\to\infty}  F(t,\lambda(t) x) = \fzero(x), \quad x \in (0,\infty), \]
where $\fzero$ is the Mittag-Leffler distribution function defined in
(\ref{eq:mittag_leffler}).
\end{enumerate}
\end{thm}

\begin{proof}
Our proof is based on the Laplace transform. We first reformulate
(\ref{eq:hyp1}) and (\ref{eq:hyp2}) in terms of $\varphi_0(s)$.
Firstly, by the well-known characterization of weak convergence by the
Laplace transform~\cite{Feller}, (\ref{eq:hyp1}) is
equivalent to the assertion that the Laplace transforms
converge pointwise, i.e., that 
\begin{equation}
\label{eq:hyp1_1}
\lim_{t \to \infty} \frac{\eta(t,s\lambda^{-1})}{\eta(t,0)} \to
\eta_*(s) : = \int_0^\infty e^{-sx} F_*(dx) , \quad s \in [0,\infty).
\end{equation}
The assumption $F_*(x)<1$ for some $x>0$ ensures that $0 < \eta_*(s)
<1$ for all $s> 0$. 
Since $\eta(t,s)=\varphi(t,\infty)-\varphi(t,s)$ and
$\varphi(t,\infty)=\eta(t,0)$, by the solution formula
(\ref{eq:const_5}) we have
\begin{equation}
\label{eq:etafrac}
\frac{\eta(t,s\lambda^{-1})}{\eta(t,0)} = 
\frac{1+\varphi_0(s\lambda^{-1})\varphi_0(\infty)^{-1}}
{1+t\varphi_0(s\lambda^{-1})}.
\end{equation}
Because $\varphi_0(0)=0$, existence of 
the limit in (\ref{eq:hyp1_1}) with $0<\eta_*(s)<1$ 
is equivalent to the existence of
\begin{equation}
\label{eq:gdef}
g(s):=\lim_{t \rightarrow \infty} t \varphi_0(s\lambda(t)^{-1})
\end{equation}
with $0<g(s)<\infty$, for all $s >0$. 

The behavior in (\ref{eq:hyp2}) may also be reformulated in terms of the 
Laplace transform. Applying Theorem~\ref{thm:tauber} and
Lemma~\ref{le:diff_phi} to the mass distribution function appearing on
the left-hand side, we find (\ref{eq:hyp2}) equivalent to
\begin{equation}
\label{eq:hyp2_1}
\varphi_0(s) \sim \frac{\Gamma(2-\rho)}{\rho} s^\rho L(\frac{1}{s}), 
\quad \text{as $s \to 0$.}
\end{equation}

Now to prove the first part of the theorem, we take $t=1, 2,\ldots$ in 
(\ref{eq:gdef}) and apply Lemma~\ref{le:rv} to conclude that 
$\varphi_0(s)$ is regularly varying at $0$ 
and $g(s)= c s^\rho$ for some $c >0$ and $\rho \ge0$.
In fact $\rho>0$ since 
$\eta_*(s) = (1+cs^\rho)^{-1}$ must satisfy $\eta_*(0)=1$.
Hence (\ref{eq:hyp2_1}) holds, and 
it remains to show that $\rho \in (0,1]$. This will follow from
complete monotonicity of the limit. 
Since $\eta(t,s)$ is completely monotone, it follows that $\eta_*(s)$
is completely monotone since it is the limit of a
sequence of completely monotone functions. 
This is possible only if $\rho \in (0,1]$, since the second derivative 
of $(1+cs^\rho)^{-1}$ is not ultimately positive if $\rho>1$ \cite{Pillai}.

Conversely, we prove the second part by showing that
(\ref{eq:hyp2_1}) implies (\ref{eq:hyp1_1}) with
$F_*=\fzero$. We define $\lambda(t)$ for
sufficiently large $t$ by 
\begin{equation}
\label{eq:time_scaling1}
 t \varphi_0(\lambda(t)^{-1}) = 1.
\end{equation}
$\lambda(t)$ is strictly
increasing because $\varphi_0(s)$ is strictly increasing. Moreover,  
$\lim_{t \rightarrow \infty} \lambda(t)=\infty$ since $\varphi(0)
=0$. Since $\varphi_0$ is regularly varying with index $\rho$ we have
\[ \lim_{t \rightarrow \infty} t \varphi_0(s\lambda^{-1}) = \lim_{t
  \rightarrow \infty} \frac{
  \varphi_0(s\lambda^{-1})}{\varphi_0(\lambda^{-1})} = s^\rho.\]
But then (\ref{eq:etafrac}) yields (\ref{eq:hyp1_1}) with
$\eta_*(s)=(1+s^\rho)^{-1}$.
\end{proof}

\begin{rem}
\label{rem:time_scaling1}
Let $\varphi_0(s) = s^\rho L(1/s)$. Then equation
(\ref{eq:time_scaling1}) shows that  
\[ \lambda(t) L(\lambda (t))^{1/\rho} = t^{1/\rho}. \]
Comparison with the time scaling $\lambda(t) = t^{1/\rho}$ for the
self-similar solution (\ref{eq:const_ansatz4}) shows that $\lambda(t)$
chosen in the proof is essentially the time scaling of the
self-similar solution, possibly modified by  a slowly varying
correction.
\end{rem}

\begin{rem}
\label{rem:domain_1}
When $\rho=1$ the condition for being attracted to the exponential
distribution is $\int_0^x y \nu_0(dy) \sim L(x)$ as $x\to \infty$. 
Thus, all solutions with initially finite mass are attracted
to the finite-mass exponential distribution, 
but it is not necessary for this 
that the initial mass be finite. It suffices
that the mass distribution function diverge sufficiently
weakly.
\end{rem}

\begin{rem}
A remaining nontrivial possibility to discuss is that a nonzero
in (\ref{eq:hyp1}) may exist 
where the function $F_*$ is a {\it defective} probability
distribution, satisfying $F_*(\infty)<1$.
If this is the case, then since $\eta_*(0^+)=F_*(\infty)<1$
it follows $g(s)=cs^\rho$ with $\rho=0$, and that 
$\varphi_0(s)\sim L(1/s)$ is slowly varying at 0. 
We cannot ensure (\ref{eq:hyp2}) in this case.
Instead we note that 
$\varphi_0(s)/s=\int_0^\infty e^{-sx}\int_x^\infty \nu_0(dy)\,dx$,
and it follows from the Tauberian theorem and the fact that 
$x\mapsto\int_x^\infty \nu_0(dy)$ is monotone (\cite{Feller}, XIII.5.4)
that the tail distribution function is slowly varying at $\infty$, with
\begin{equation}\label{eq:tail}
\nu_0((x,\infty)) = \int_x^\infty \nu_0(dx) \sim L(x).
\end{equation}
Conversely, if (\ref{eq:tail}) holds, 
then $\varphi_0(s)$ is a function slowly varying at 0
that strictly increases. For any $c\in(0,\infty)$, 
we can choose $\lambda(t)$ strictly
increasing such that $t\varphi(\lambda(t)^{-1}) = c$.
Then it follows that (\ref{eq:hyp1_1}) holds with
$\eta_*(s)=(1+c)^{-1}$ for $s>0$, 
so (\ref{eq:hyp1}) holds with the defective distribution function
$F_*(x)=(1+c)^{-1}$.  This means that under such scalings,
an arbitrary fraction of the particle sizes concentrate
at 0 and the rest escapes to infinity.
\end{rem}

\section{Scaling solutions for the additive kernel}

\subsection{A one-parameter family of solutions}
Golovin found an exact solution to Smoluchowski's equations
with monodisperse initial condition for $K=x+y$. 
One may take limits in his solution to
obtain the scaling solution~\cite{Aldous}
\begin{equation}
\label{eq:golovin}
n(t,x) = \frac{1}{\sqrt{2\pi}}x^{-3/2} e^{-t} \exp({-e^{-2t}x/2})
\end{equation} 
This solution has sometimes been criticized as unphysical, 
since the number of clusters is infinite. 
However, recently Deaconu and Tanr\'{e} proved a result equivalent 
to weak convergence to this solution, 
under restrictive assumptions on the initial data (the existence
of all moments and their domination by the moments of a Gaussian
random variable)~\cite{DT}.  

We  will consider only solutions of finite mass, normalized to 1. 
The solution (\ref{eq:golovin}) is but one of a one-parameter family
of solutions (independent of the trivial scaling $c^2n(t,cx)$). 
For each $\rho \in (0,1]$, in this section we derive 
finite-mass scaling solutions in the following form,
with $\beta=\rho/(1+\rho)$:
\begin{equation}\label{eq:nform}
n(t,x) = e^{-2t/\beta} n_\rho(e^{-t/\beta}x),
\end{equation}
where 
\begin{equation}
\label{eq:l6}
 n_\rho(x) = \frac1\pi \sum_{k=1}^\infty
 \frac{(-1)^{k-1}x^{k\beta-2}}{k!}
\Gamma(1+k-k\beta){\sin \pi k\beta}.
\end{equation}
The associated mass distribution function is given by
\begin{equation}\label{eq:Mdef}
M_\rho(x) = \int_0^xyn_\rho(y)\,dy = 
 \sum_{k=1}^\infty \frac{(-1)^{k-1}x^{k\beta}}{k!}
\Gamma(1+k-k\beta)\frac{\sin \pi k\beta}{\pi k\beta}.
\end{equation}

\begin{rem}\label{rem:stable}
It is an interesting fact that these scaling solutions
are related by a nonlinear scaling to the
{\em \Levy\ stable laws} in probability theory.
Feller (\cite{Feller}, XVII.7) gives the formula
\begin{equation}\label{eq:plaw}
p(x;\alpha,\gamma)=
\frac1{\pi x}\sum_{k=1}^\infty
\frac{(-x)^k}{k!}\Gamma(1+k/\alpha)\sin\frac{k\pi}{2\alpha}(\gamma-\alpha)
\end{equation}
for a family of stable densities, for $1<\alpha<2$, $|\gamma|\le2-\alpha$.
Taking $\alpha=(1-\beta)^{-1}$, $\gamma=2-\alpha$, we find 
that the mass density from (\ref{eq:l6}) satisfies
\begin{equation}\label{eq:nlaw}
x n_\rho(x) = x^{\beta-1}p(x^\beta;1+\rho,1-\rho).
\end{equation}
These remarkable self-similar solutions were first 
discovered by Bertoin~\cite{Bertoin} (then independently by us).
Bertoin's derivation explains the nonlinear
rescaling formula in terms of a scaling property of \Levy\ stable
processes, and he writes the self-similar solution in the form
\begin{equation}
\label{eq:nlaw2}
n(t,x) = e^{-t} x^{\beta-2} p(e^{-t}x^\beta;1+\rho,1-\rho).
\end{equation}
It is quite remarkable that there are {\em two\/} scaling limits associated to
this solution. One of them is more transparent in (\ref{eq:nlaw2}): we
find that $e^t n(t,x) dx$ converges vaguely towards the measure
$x^{\beta-2} dx$. This is the scaling limit alluded to in Corollary 1 of
Bertoin's article~\cite{Bertoin}, and it suffices to uniquely identify the
self-similar solution in the class of eternal solutions. 
On the other hand, (\ref{eq:nform}), reflects
more clearly the self-similar nature of the solution relative
to the mean cluster size $e^{t/\beta}$.

Note that the stable densities are defined on the whole line 
$(-\infty,\infty)$.  We obtain total mass 1 on $(0,\infty)$ through 
the nonlinear rescaling. Moreover,
if $F(x;\alpha,\gamma)$ denotes the distribution function for the stable
law with density $p(x;\alpha,\gamma)$, then the tail of the mass
distribution corresponds to this through
\begin{equation}\label{eq:Mtail}
1-M_\rho(x) = \beta^{-1}(1-F(x^\beta;1+\rho,1-\rho)).
\end{equation}
\end{rem}

The total number of clusters diverges for all the solutions in
(\ref{eq:nform}).
This is caused by the predominance of small clusters, and it may be
desingularized by working with the variable $\varphi$ 
introduced in Section~\ref{sec:renorm}. 
The scaling solutions above are given implicitly in terms of
$\varphi_\rho=\varphi_\rho(s)$ satisfying 
\begin{equation}
\label{eq:add_scaling1}
s = \varphi_\rho + \varphi_\rho^{1+\rho}.
\end{equation}
Since $u_\rho(s) = \partial_s \varphi_\rho$ is the Laplace transform
of the mass distribution, differentiating (\ref{eq:add_scaling1}) we have
\begin{equation}
\label{eq:add_scaling}
u_\rho = \frac{1}{1 + (1+\rho) \varphi_\rho^\rho},
\end{equation}
which exhibits a connection to the Mittag-Leffler distribution (see
(\ref{eq:ML_trans})). We use this to show below that the mass
distribution is
infinitely divisible. When $\rho=1$, equation (\ref{eq:add_scaling1})
is a quadratic equation with two solutions, one of which is $\varphi(s) =
\sqrt{1+4s}- 1$, which corresponds to (\ref{eq:golovin}). 
For $\rho \in (0,1)$  we will solve (\ref{eq:add_scaling1}) by Laplace's
inversion formula as an infinite series, to obtain (\ref{eq:l6}).

\subsection{Scaling solutions}
We may derive mass-preserving scaling solutions to (\ref{eq:evol-add})
as follows. Let $\lambda(t)$ be a rescaling to be determined, and let
$\xi =s\lambda$. We substitute the ansatz
$\varphi(t,s)=\lambda^{-1}\varphi_\rho(s\lambda) = \lambda^{-1}
\varphi_\rho(\xi)$, in (\ref{eq:evol-add}) to obtain
\begin{equation}
\label{eq:add_sc1}
- \frac{\dot{\lambda}}{\lambda} (\varphi_\rho - \xi
\partial_{\xi} \varphi_\rho) - \varphi_\rho
\partial_{\xi} \varphi_\rho = -\varphi_\rho.  
\end{equation}
We separate variables by letting 
${\dot{\lambda}}/{\lambda} = a$ or $ \lambda = c_1 e^{at}$. Then by
(\ref{eq:add_sc1})  
\begin{equation}
\label{eq:add_sc3}
 (a\xi -\varphi_\rho) \partial_\xi \varphi_\rho + (1-a) \varphi_\rho=0.
\end{equation}
Equation (\ref{eq:add_sc3}) is not separable, but it may be solved
implicitly by rewriting it as the linear equation
\begin{equation}
\label{eq:add_sc4}
\frac{d\xi}{d\varphi_\rho} - \frac{a}{a-1} \frac{\xi}{\varphi_\rho} =
\frac{1}{1-a}. 
\end{equation}
Put $\rho = (a-1)^{-1}$ so $a=(1+\rho)/\rho=1/\beta$.
Integrating, we find a family of nontrivial solutions
determined by
\begin{equation}
\label{eq:add_sc5}
  \xi = \varphi_\rho + c_2 \varphi_\rho^{1+\rho}, \quad c_2>0.
\end{equation}
The range of admissible $\rho$ is narrowed
by requiring that $\lim_{\xi \rightarrow 0} 
\varphi_\rho/\xi=1$  (finite mass) which implies $\rho >0$.
Without loss of generality we may take $c_2=1$, since we can
recover all other solutions by a trivial scaling.

We now show that $\rho >1$ is inadmissible.
Let $U(\xi)= \varphi(\xi)/\xi$. Integrating (\ref{eq:renorm_6}) by parts we see
that $U(\xi) = \int_0^\infty e^{-\xi x} N(x) dx$ where $N(x) =
\nu((x,\infty))$ is the tail distribution. In particular, $U(0)=1$ and $U$ is
completely monotone. Dividing (\ref{eq:add_sc5}) by $\xi$ and
differentiating we see that
\[ U'(\xi) = \frac{ -c \xi^{\rho-1} U^{1+\rho}}
{1+c(\rho+1) \xi^\rho U^\rho} \to 0 \quad\mbox{as $\xi\to0$,} \]
which is impossible if $U$ is completely monotone. Thus, the
admissible range of non-trivial solutions is restricted to $\rho \in
(0,1]$. 

\subsection{Series expansion and asymptotics of scaling solutions}
The asymptotic properties of the scaling solutions for $\rho \in
(0,1)$ may be rigorously obtained from Theorem~\ref{thm:tauber}.
By (\ref{eq:add_scaling1}),
$\varphi_\rho'' \sim \rho(\rho +1) s^{\rho-1}$ as  $s \rightarrow 0$. 
But, $\varphi_\rho'' = \int_0^\infty e^{-sx} x^2 n_\rho(x) dx$, and
Theorem~\ref{thm:tauber} implies that
\begin{equation}
\label{eq:tauber_2}
\int_0^x y^2 n_\rho(y) dy \sim \frac{
  \rho (\rho+1)}{\Gamma(2-\rho)}x^{1-\rho}, 
  \quad\text{as $x \rightarrow \infty.$}
\end{equation}
Thus the second moment is finite only for $\rho=1$. For $0<\rho<1$
the mass distribution has fat tails.
Equation (\ref{eq:tauber_2}) is a weak version of 
the pointwise behavior
\begin{equation}
\label{eq:tauber_3}
n_\rho(x) \sim \frac{\rho +1}{|\Gamma(-\rho)|}x^{-(2+\rho)} \quad
\text{as $x \rightarrow \infty$},\  \rho \in (0,1),
\end{equation}
which follows from (\ref{eq:nlaw}) due to the known power-law 
asymptotics of the stable densities~\cite{Bergstrom}.

The behavior as $x \rightarrow 0$ is described completely by the
series (\ref{eq:l6}), derived as follows.  
We rewrite (\ref{eq:add_scaling1}) in terms
of $U =\varphi_\rho/s$ and $\beta=\rho/(1+\rho)$ as  
\begin{equation}
\nonumber
U(s) = s^{-\beta} (1 - U)^{1-\beta}.
\end{equation}
We solve for $U$ using Lagrange's inversion formula 
(see e.g.~\cite[6.3]{Kingman} for a similar calculation), obtaining
\begin{equation}
\nonumber
U(s) = \sum_{k=1}^\infty \frac{s^{-k\beta}}{k!}
\frac{d^{k-1}}{dx^{k-1}} \left(F(x)\right)^k \vert_{x=0},
\quad\mbox{with}\quad F(x) = (1-x)^{1-\beta}. 
\end{equation}
We evaluate the derivatives and find that
\begin{equation}
\label{eq:l4}
\D_s\varphi_\rho(s) = \partial_s(sU) = 
\sum_{k=1}^\infty \frac{s^{-k\beta}}{k!}
(-1)^{k-1}\prod_{j=1}^k (j-k\beta). 
\end{equation}
This is the Laplace transform of the mass distribution function
given through term-by-term Laplace inversion as
\begin{equation}
M_\rho(x) =  \sum_{k=1}^\infty
\frac {(-1)^{k-1}} {k!}\frac {x^{k\beta}} {\Gamma(1+k\beta)}
\prod_{j=1}^k (j-k\beta) 
\label{eq:Mrho}
\end{equation}
We then deduce (\ref{eq:Mdef}) using $y=k\beta$ in the gamma-function 
identity (\ref{eq:Gamma1}).
By differentiating (\ref{eq:Mdef}) we obtain the number density in 
(\ref{eq:l6}).

It is straightforward to check that when $\rho=1$, the even
terms vanish and (\ref{eq:l6})
reduces to the function $(4\pi)^{-1/2}x^{-3/2}e^{-x/4}$ which is a
scaled version of (\ref{eq:golovin}). 
Correspondingly, $M_1(x)=\mathop{\rm erf}(\frac12\sqrt{x})$.
One may also check that the series
solution above is absolutely convergent for $x \in (0,\infty)$.
Thus, $n_\rho$ is analytic. 

\section{Weak convergence for the additive kernel}
We let $\nu_t$ be a solution of Smoluchowski's equation with
kernel $K=x+y$ given initial data $\nu_0$ with finite mass normalized
to $\int_0^\infty x\,\nu_0(dx)=1$.
Then at all times the mass distribution is a probability distribution,
with distribution function denoted
\[ M(t,x) = \int_0^x y \nu_t(dy). \]
It follows from the explicit solution (\ref{eq:add_soln3}) that
the Laplace transform of the mass distribution satisfies
$u(t,0) =1$ for all $t \geq 0$, and  
$\lim_{t \rightarrow \infty} u(t,s) = 0$ for $s >0$.
This phenomenon of concentration is equivalent to the assertion that
asymptotically all the
mass escapes to infinity. As earlier, we
may hope that suitable rescaling in $s$ will give convergence to a
nontrivial limit. Precisely, we have the following characterization.
\begin{thm}
\label{thm:add}
\begin{enumerate}
\item Suppose there is a rescaling function 
$\lambda(t) \rightarrow \infty$ as $t\to\infty$ and a nontrivial probability
distribution function $M_*(x)$ such that 
\begin{equation}
\label{eq:add5}
\lim_{t \rightarrow \infty} M(t,\lambda(t)x) = M_*(x) 
\end{equation}
at all points of continuity of $M_*$.
Then there exists $\rho \in (0,1]$ 
and a function $L$ slowly varying at infinity such that 
\begin{equation}
\label{eq:add_M2}
\int_0^x y^2 \num_0(dy) \sim {x^{1-\rho} L(x)}
\quad\text{as $ x \to \infty.$} 
\end{equation}
\item Conversely, assume that there exists $\rho \in (0,1]$ and a
function $L$ slowly varying at infinity such that (\ref{eq:add_M2})
  holds. Then there is a strictly increasing rescaling $\lambda(t) \to
  \infty$ such that
\[ \lim_{t\to\infty} M(t,\lambda(t) x) = M_\rho(x), \quad 0 \leq x < \infty,\]
where $M_\rho$ from (\ref{eq:Mdef})
is the mass distribution function for a scaling solution.
\end{enumerate}
\end{thm}
\begin{proof}
We will prove the theorem after reformulating $(1)$ and $(2)$ as
equivalent assertions using the Laplace transform. Firstly, the weak
convergence of the mass distribution $M(t,\lambda x)$ is equivalent to
the pointwise convergence of its Laplace transform
\begin{equation}
\label{eq:add_reform1}
\lim_{t \to \infty} u(t,s\lambda^{-1}) = u_*(s) :=\int_0^\infty
e^{-sx} M_*(dx), \quad 0\leq s< \infty.
\end{equation}
The assumption $M_*(x) <1$ for some $x >0$ ensures that $0 < u_*(s)
<1$ for $s>0$. Secondly, (\ref{eq:add_M2}) is equivalent to 
$-\partial_s u \sim s^{\rho-1} L(1/s)\Gamma(2-\rho)$ as $s \to 0$ by
Theorem~\ref{thm:tauber}. Since $\rho \in (0,1]$,  by
Lemma~\ref{le:diff_phi} this is equivalent to 
\begin{equation}
\label{eq:add_reform2}
1-u_0(s) \sim s^\rho L\left(\frac{1}{s}\right) \frac{\Gamma(2-\rho)}{\rho}
\quad \text{as $s \to 0.$}
\end{equation}

We prove the first part of the theorem by showing that
(\ref{eq:add_reform1}) implies (\ref{eq:add_reform2}). Since 
$u_*(s)$ is a limit of completely monotone functions, it is itself
completely monotone. Moreover, since $u(t,s) = \partial_s \varphi$ we
also have the convergence
\begin{equation}
\label{eq:phi_conv}
 \lim_{t \to \infty} \lambda \varphi(t, s\lambda^{-1}) =
\varphi_*(s) = \int_0^s u_*(s') ds'. 
\end{equation}
Clearly, $\varphi_*$ is strictly increasing. 

In what follows, we consider $\sigma(t,s)$ defined by replacing $s$ with
$s/\lambda(t)$ in (\ref{eq:add_soln2}), i.e., by
\begin{equation}
\label{eq:add6}
\sigma - \varphi_0(\sigma)= s\lambda^{-1} - \varphi(t, s\lambda^{-1}). 
\end{equation}
It then follows from (\ref{eq:add6}) that as $t\to\infty$ with $s$
fixed, we have $\sigma \to 0$, and
\begin{equation}
\label{eq:add_pr1}
\lim_{t \to \infty} \lambda(\sigma - \varphi_0(\sigma)) = s - \varphi_*(s). 
\end{equation}
From (\ref{eq:add_soln1}) and (\ref{eq:phi_conv}) we then have 
\begin{equation}
\label{eq:add_pr2}
\lim_{t \to
  \infty} \lambda  e^{-t} \varphi_0(\sigma) =  \lim_{t  \to \infty}
  \lambda \varphi (t, s\lambda^{-1}) =  
  \varphi_*(s). 
\end{equation}
Replacing $s$ by $s\lambda^{-1}$ in the exact solution
(\ref{eq:add_soln3}) we also have
\[ \lim_{t \to \infty} \frac{u_0(\sigma)}{e^t(1-u_0(\sigma)) +
  u_0(\sigma)} = u_*(s).\]
Since $\sigma \to 0$ and $u_0(0)=1$ we deduce that
\begin{equation}
\label{eq:add_pr3}
\lim_{t \to \infty} e^t (1-u_0(\sigma)) = \frac{1-u_*(s)}{u_*(s)}.
\end{equation}
We now show that $\sigma  \to 0$ at the rate  $a(t):= e^t \lambda(t)^{-1}$ 
($\lim_{t \to \infty} a(t)=0$ by (\ref{eq:add_pr2})). We may rewrite
(\ref{eq:add6}) as 
\begin{eqnarray}
\nonumber
\sigma &=&  a(t) \left[ \varphi_*(s) + s e^{-t} + \lambda
    e^{-t}\varphi_0(\sigma)(1-e^{-t}) - \varphi_*(s) \right] \\
\nonumber
& =& a(t)[\varphi_*(s) + r(t,s)] ,
\end{eqnarray}
where the error term $r(t,s) \to 0$ by (\ref{eq:add_pr2}). Therefore,
$\sigma$ is asymptotically a scaling of $\varphi_*(s)$.  We now 
claim that for all $s>0$, 
\begin{equation}
\label{Vlim}
\lim_{t\to\infty} e^t (1-u_0(a(t)\varphi_*(s))) =
\frac{1-u_*(s)}{u_*(s)}  .
\end{equation}
Fix $s >0$, and let $\delta>0$ be sufficiently small. Since
$\varphi_*$ is strictly increasing, 
for sufficiently large $t$ (depending on $s$ and $\delta$) we have
\[
\varphi_*(s-\delta)+r(t,s-\delta) < \varphi_*(s) <
\varphi_*(s+\delta)+r(t,s+\delta), 
\]
whence
\[1-u_0(\sigma(t,s-\delta)) < 1-u_0(a(t)\varphi_*) < 
1-u_0(\sigma(t,s+\delta)).\]
Multiply by $e^t$ and take $t\to\infty$, then $\delta\to0$. 
The claim (\ref{Vlim}) then follows from (\ref{eq:add_pr3}).

It follows directly from (\ref{Vlim}) and Lemma~\ref{le:rv} that
$1-u_0$ is regularly varying at 0 with some exponent
$\rho\in\mathbb{R}$,
and the limit in (\ref{Vlim}) has the form $c \varphi_*^\rho$ for some
positive constant $c$.
Clearly $0< \rho\le1$ since $u_0$ is bounded and completely
monotone and $u_*(0)=1$ by the hypothesis that $M_*$ is a
probability distribution.  This finishes the proof of the first part.
Note furthermore that (\ref{eq:add_scaling}) holds after scaling $s$.

We prove the converse statement by
showing that (\ref{eq:add_reform2}) implies (\ref{eq:add_reform1}) with
$u_*=u_\rho$.  By the explicit
solution formula (\ref{eq:add_soln3}), it suffices to show that
as $t\to\infty$ we have
\begin{equation}\label{eq:u0lim}
e^t (1-u_0(\sigma(t,s\lambda^{-1}))) = 
 {u(t,s\lambda^{-1})^{-1}}-1 \to (1+\rho)\varphi_\rho(s)^\rho
\end{equation}
for all $s>0$.  We write $a(t) = e^t \lambda(t)^{-1}$ as earlier. 
We choose $\lambda(t)$ to satisfy $\lambda(0)=0$ and  
\begin{equation}
\label{eq:add_lambda}
 e^t(1 -u_0(a(t))) = 1+\rho, \quad t>0.
\end{equation}
Then $a(t)\to0$, $\lambda(t)$ is strictly increasing and 
$\lambda(t) \rightarrow \infty$.  
It follows from (\ref{eq:add_reform2})  that for fixed
$\varphi_*>0$, as $t\to\infty$ we have 
\begin{equation}
\label{eq:add_lambda2}
  e^t(1 -u_0(a(t)\varphi_*)) = (1+\rho) 
 \frac{1-u_0(a(t) \varphi_*)}{1-u_0(a(t))} \to (1+\rho) \varphi_*^\rho.
\end{equation}
For fixed $\varphi_* >0$ we define  
$s(t,\varphi_*)$ as the value of $s$ determined from (\ref{eq:add6})
using $\sigma = a(t)\varphi_*$. Then it follows that for all
$\varphi_*>0$,
\begin{equation}\label{eq:ulim}
\lim_{t\to\infty} u(t,s(t,\varphi_*)\lambda^{-1})^{-1} -1 =
{(1+\rho)\varphi_*^\rho}.
\end{equation}
Using (\ref{eq:add_soln1}) with (\ref{eq:add6}) we have
\begin{eqnarray*}
s(t,\varphi_*) &=& \lambda(a\varphi_*-\varphi_0(a\varphi_*))
+ \lambda e^{-t} \varphi_0(a\varphi_*)  \\
&=& (1+\rho)\frac{ a\varphi_*-\varphi_0(a\varphi_*)}{a(1-u_0(a))}
+\frac1a \varphi_0(a\varphi_*).
\end{eqnarray*}
Since $1-u_0 \sim s^\rho L(1/s)$, the proof of Lemma~\ref{le:diff_phi} shows
that $s-\varphi_0(s) \sim (1+\rho)^{-1} s^{\rho+1}L(1/s)$. Therefore,
as $t\to\infty$ we have
\begin{equation}
s(t,\varphi_*) \sim \frac{(a\varphi_*)^{1+\rho} L(1/a\varphi_*)}
{a^{1+\rho}L(1/a)} +\varphi_* \to 
\varphi_*^{1+\rho}+\varphi_*.
\end{equation}
Now to prove (\ref{eq:u0lim}), fix $s_0>0$. 
Then there is a unique $\varphi_*=\varphi_\rho(s_0)>0$ so that $s_0 =
\varphi_*(\varphi_*^\rho+1)$.  
Since $\varphi_*\mapsto s(t,\varphi_*)$ is strictly increasing in
$\varphi_*$ for all $t$,  
by substituting $\varphi_*  \pm\delta$ for $\varphi_*$ in (\ref{eq:ulim})
we easily deduce (\ref{eq:u0lim}).
\end{proof}

\begin{rem}
When $1-u_0 = s^{\rho}$, the choice of time scaling $\lambda(t)$ in
(\ref{eq:add_lambda}) gives $\lambda(t) = e^{(1+1/\rho)t}$ in
accordance with (\ref{eq:add_sc3}). More
generally, when $1-u_0 = s^\rho L(1/s)$ the rescaling $\lambda(t)$ is
modified by a slowly varying correction. 
The choice of time scale when $\rho=1$ and the second moment
is finite deserves special comment. In this case we find $\lambda(t)=
e^{2t}$. In the applied literature it is common to define mean cluster
size as a ratio of moments. Let $\mmt_k(t)= \int_0^\infty x^k
\nu_t(dx)$. It is clear that any ratio of the form $\mmt_{k+1}(t)/\mmt_k(t)$
has the dimensions of length, and two distinct, but 
natural definitions of mean cluster size are (see~\cite{DE1})
\[ c_1(t) = \frac{\mmt_1(t)}{\mmt_0(t)} \quad\mbox{and}\quad 
c_2(t) = \frac{\mmt_2(t)}{\mmt_1(t)}.\]
For Golovin's solution the initial data are monodisperse, and
$\mmt_0(0)=\mmt_1(0)=\mmt_2(0)=1$. It is easy to calculate explicitly that
$c_1(t)=e^t$, but $c_2(t) =e^{2t}$. Thus the two notions of cluster
size differ, and only $e^{2t}$ is the correct scaling for convergence
to self-similar form. More generally, $e^{2t}$ is the only choice of
time scaling that fixes {\em both\/} $\mmt_1$ and $\mmt_2$ as required for
convergence to self-similar form.
\end{rem}

\begin{rem}
Theorem~\ref{thm:add} with $\rho=1$ shows that the mass distribution 
is attracted to the classical self-similar solution for
all initial data $\nu_0$ with finite first and second moment. 
But for this behavior it is not necessary for the initial data to
have finite second moment. It suffices that it diverge sufficiently
weakly, with $\int_0^x y^2\nu_0(dy) \sim L(x)$ slowly varying at
infinity.
\end{rem}

\begin{rem}
A remaining nontrivial possibility is that a nonzero
limit in (\ref{eq:add5}) may exist 
with $M_*$ defective, satisfying $M_*(\infty)<1$.
If this is true, then most of the proof of the first part
of the theorem carries through. 
The limit in (\ref{Vlim}) must have the form $c\varphi_*^\rho$ with 
$c>0$, but we must have $\rho=0$, since $u_*(0)<1$. 
Moreover it follows $1-u_0(s)\sim L(1/s)$ is slowly varying.
We do not obtain (\ref{eq:add_M2}) in this case. Instead, we note
\begin{equation}
\frac{1-u_0(s)}{s} = \int_0^\infty e^{-sx}\int_x^\infty y\,\nu_0(dy)\,dx.
\end{equation}
As for the constant kernel, it follows from the Tauberian theorem 
and monotonicity that the tail of the mass distribution is slowly
varying at infinity, with
\begin{equation}\label{eq:Mtail2}
\int_x^\infty y\,\nu_0(dy)\sim L(x) .
\end{equation}
In the converse direction, if (\ref{eq:Mtail2}) holds, then
$1-u_0(s)$ is a strictly increasing function that is slowly varying at
0. For any $c\in(0,\infty)$ we can choose $\lambda(t)$ strictly
increasing so that $e^t(1-u_0(a(t)))= c$.
Then it follows as in the proof of the second part of the theorem
that (\ref{eq:add_reform1}) holds with
$u_*(s)=(1+c)^{-1}$ for $s>0$, 
so (\ref{eq:add5}) holds with the defective distribution function
$M_*(x)=(1+c)^{-1}$.  This means that under such scalings,
an arbitrary fraction of the mass concentrates
at 0 and the rest escapes to infinity.
\end{rem}

We conclude this section with a useful observation about the
self-similar solutions.
\begin{thm}
\label{thm:id_add}
For each $\rho\in(0,1]$, the probability distribution $M_\rho$ is infinitely
divisible.
\end{thm}
\begin{proof}
It suffices to show that the Laplace transform $u_\rho=
e^{-\psi_\rho}$ where $\psi_\rho(0)=0$, and $\psi_\rho$ has completely
monotone derivative~\cite[XIII.7.1]{Feller}. By (\ref{eq:add_scaling}),
$\psi_\rho = \log(1+ (1+\rho) \varphi_\rho^\rho)$. Clearly,
$\psi_\rho(0)=0$.
Moreover,
\[ \partial_s \psi_\rho = \frac{(1+\rho) \rho \varphi_\rho^{\rho-1}
  u_\rho}{1+(1+\rho)\varphi_\rho^\rho}. \]
By Theorem~\ref{thm:add} $u_\rho$ is completely monotone. The other
factor can be written as a composition with the Mittag-Leffler
distribution 
\[ \frac{\varphi_\rho^{\rho-1}}{1+\varphi_\rho^\rho} =
\frac{s^{\rho-1}}{1+s^\rho} \circ 
\varphi_\rho. \]
The function $s^{\rho-1}$ is completely monotone, as is
$(1+s^\rho)^{-1}$. Thus, their product is completely monotone. 
Thus, the composed function above
is completely monotone since it is the composition of a
completely monotone function with a function that has a completely
monotone derivative~\cite[XIII.4.2]{Feller}. Finally, $\partial_s
\psi_\rho$ is 
the product of two completely monotone functions, and is hence
completely monotone.
\end{proof}

\section{Approach to self-similar gelation for the multiplicative kernel}
\subsection{McLeod's solution}
McLeod found the following explicit solution to the discrete
Smoluchowski equation (\ref{eq:smol1}) for $K=xy$ and
monodisperse initial data $\nu_0=\delta(x-1)$~\cite{Aldous,McL}:
\begin{equation}
\label{eq:McL}
 \nu_t = \sum_{k=1}^\infty n_k(t)\delta(x-k), \quad
 n_k(t)= \frac{t^{k-1}k^{k-2}}{k!e^{tk}}.
\end{equation}
A beautiful probabilistic interpretation of this solution in terms of
a Poisson-Galton-Watson branching process may be found in~\cite{Aldous}.
The solution is valid only for $0\le t<1$. When $t=1$, $n_k(t)$ only
has algebraic decay, and the second moment $\mmt_2(t) =
\infty$. Moreover, mass can no longer be conserved for $t>1$. 
At a microscopic level, this is commonly ascribed
to the formation of a cluster of infinite mass (the gel).  

The formal scaling limit of (\ref{eq:McL}) is obtained by considering
the large $k$ limit as $t \rightarrow 1$. By Stirling's approximation
$k! \sim \sqrt{2\pi k} e^{-k} k^k$, as $t \to 1$ we find
\begin{eqnarray*}
n_k(t) &\sim& \frac{1}{\sqrt{2\pi}}k^{-5/2}  e^{k \left( 1-t+
      \log t \right) } 
\\
& \sim & \frac{1}{\sqrt{2\pi}} k^{-5/2}\exp\left(-k((1-t)^2/2 + (1-t)^3/3 +
  \ldots)\right) .
\end{eqnarray*}
Let $x = k(1-t)^2$ and consider the limit $k\to \infty, t \to 1$ such
that $x$ is held fixed. Thus, we find
\[ \lim_{t \to 1, k \to \infty} (1-t)^{-5} n_k(t) =
\frac{1}{\sqrt{2\pi}}x^{-5/2} e^{-x/2}. \] 
This shows convergence of the discrete solution to the scaling
solution~\cite{Aldous}
\begin{equation}
\label{eq:McL2}
 n(t,x) = \frac{1}{\sqrt{2 \pi}} x^{-5/2} e^{-(1-t)^2 x/2}, \quad x \in
 (0,\infty),\ t \in(-\infty,1).
\end{equation}
We will see below that this scaling solution emerges coherently from
the scaling solutions to the additive kernel,
and is just one of a one-parameter family of scaling solutions.

\subsection{Scaling solutions and weak convergence}
The scaling solutions for the multiplicative kernel can be obtained by
our knowledge of the scaling solutions to the additive kernel, via 
a general relation between solutions for the two kernels.
Recall from Section~\ref{subsec:mult} that the initial data are
normalized so the initial second moment $\mmt_2(0)=1$. 
Then $\mmt_2(t)=(1-t)^{-1}$, and 
the gelation time $\Tgel=1$. 
The second-moment probability distribution function
\begin{equation}
\label{eq:sec_moment}
 V(t,x) = \int_0^x y^2 \nu_t(dy) \left\slash
 \int_0^\infty y^2 \nu_t(dy) \right.
\end{equation}
is the analogue of $M(t,x)$ for the additive kernel. From
(\ref{eq:defn_psi}) we see that the Laplace
transform of $V(t,x)$ is $(1-t)\partial_s \psi(t,s)$. We differentiate
equation (\ref{eq:mult4}) with respect to $s$ to obtain 
\begin{equation}
(1-t) \partial_s \psi(t,s) = 
\partial_s \varphi(-\log(1-t),s) = u(\tau,s), 
\end{equation}
where $\tau(t):= \log(1-t)^{-1}$. By consequence, 
\begin{equation}
V(t,x)=\tilde M(\tau,x),
\end{equation}
where $\tilde M(\tau,x)$ 
is the mass distribution function for the corresponding solution with 
additive kernel. For solutions with densities $n(t,x)$ and
$\tilde{n}(\tau,x)$ for the multiplicative and additive kernels respectively,
this means
\begin{equation}
x^2 n(t,x) = (1-t)^{-1}x \tilde{n}(\tau,x).
\end{equation}

From this relation we obtain the scaling solutions for the
multiplicative kernel as described in the introduction.
Explicitly,
\begin{equation}\label{eq:mult_scale}
n(t,x) = (1-t)^{-1+3/\beta}n_\rho(x(1-t)^{1/\beta}),
\end{equation}
where $\beta=\rho/(1+\rho)$ and 
\begin{equation}
 n_\rho(x) = \frac1\pi \sum_{k=1}^\infty
 \frac{(-1)^{k-1}x^{k\beta-3}}{k!}
\Gamma(1+k-k\beta){\sin \pi k\beta}.
\end{equation}
Notice that these scaling solutions do {\em not\/}
preserve mass --- in fact, all of them have infinite mass! 
Instead, they have a finite second moment for $t<1$, which blows up as
$t\to1$.  For $0<\rho<1$ the third moment is infinite.
When $\rho=1$ the scaling solution reduces to the exponentially
decaying solution in (\ref{eq:McL2}) after a trivial scaling. 
Finally, we note that though we have assumed $t \in [0,1)$,
these solutions are well-defined for $t \in (-\infty,1)$. 

Theorem~\ref{thm:add} characterizes the convergence of 
$\tilde M(\tau,\lambda x)$ and it is easy to adapt to 
characterize convergence to
self-similar form approaching the gelation time. 

\begin{thm}
\label{thm:mult}
\begin{enumerate}
\item Suppose there is a rescaling function 
$\lambda(t) \rightarrow \infty$ as $t\to 1$ and a 
nontrivial probability distribution function $V_*(x)$ such that 
\begin{equation}
\label{eq:mult5}
\lim_{t \rightarrow 1}  V(t,\lambda(t)x) = V_*(x),
\end{equation}
at all points of continuity of $V_*$.
Then there exists $\rho \in (0,1]$ 
and a function $L$ slowly varying at infinity such that
\begin{equation}
\label{eq:mult_M2}
\int_0^x y^3 \num_0(dy) \sim {x^{1-\rho} L(x)}
\quad\text{as $ x \to \infty$.} 
\end{equation}
\item Conversely, assume that there exists $\rho \in (0,1]$ 
and a function $L$ slowly varying at infinity such that (\ref{eq:mult5})
holds. Then there is a strictly increasing rescaling $\lambda(t) \to
  \infty$ such that
\[ \lim_{t\to1} V(t,\lambda(t) x) = V_\rho(x), 
\quad 0 \leq x < \infty,\]
where $V_\rho$ is the second moment distribution function for
a scaling solution, given by $V_\rho=M_\rho$ from (\ref{eq:Mdef}).
\end{enumerate}
\end{thm}

It is worth pointing out explicitly that the domain of attraction of
the scaling solution in (\ref{eq:McL2}) includes all initial data with
finite second and third moments, as well as data whose third moment
diverges sufficiently weakly (the case $\rho=1$ above).
Each of the infinite-mass self-similar solutions, however, attracts finite-mass
solutions whose third moment diverges at the appropriate rate
detailed in the theorem.

The behavior of the rescaling function $\lambda(t)$
and the characterization of possibly defective limits
can be easily deduced from the corresponding results for the additive
case that appear in the remarks following Theorem~\ref{thm:add}.

\section*{Acknowledgements}
The authors thank the Max Planck Institute for Mathematics in the
Sciences, Leipzig,  
for hospitality during part of this work. G.M. thanks Timo
Sepp\"{a}l\"{a}inen for his help during early stages of this work. 
This material is based upon work supported by the National Science
Foundation under grant nos.\ DMS 00-72609 and DMS 03-05985.

\vspace{-6pt}
\bibliographystyle{siam}
\bibliography{smol}

\begin{thebibliography}{10}

\bibitem{Abramowitz}
{\sc M.~Abramowitz and I.~A. Stegun}, {\em Handbook of mathematical functions
  with formulas, graphs, and mathematical tables}, vol.~55 of National Bureau
  of Standards Applied Mathematics Series, For sale by the Superintendent of
  Documents, U.S. Government Printing Office, Washington, D.C., 1964.

\bibitem{Aldous2}
{\sc D.~Aldous and J.~Pitman}, {\em Tree-valued {M}arkov chains derived from
  {G}alton-{W}atson processes}, Ann. Inst. H. Poincar\'e Probab. Statist., 34
  (1998), pp.~637--686.

\bibitem{Aldous-Pitman}
\leavevmode\vrule height 2pt depth -1.6pt width 23pt, {\em Inhomogeneous
  continuum random trees and the entrance boundary of the additive coalescent},
  Probab. Theory Related Fields, 118 (2000), pp.~455--482.

\bibitem{Aldous}
{\sc D.~J. Aldous}, {\em Deterministic and stochastic models for coalescence
  (aggregation and coagulation): a review of the mean-field theory for
  probabilists}, Bernoulli, 5 (1999), pp.~3--48.

\bibitem{Bergstrom}
{\sc H.~Bergstr{\"o}m}, {\em On some expansions of stable distribution
  functions}, Ark. Mat., 2 (1952), pp.~375--378.

\bibitem{Bertoin3}
{\sc J.~Bertoin}, {\em The inviscid {B}urgers equation with {B}rownian initial
  velocity}, Comm. Math. Phys., 193 (1998), pp.~397--406.

\bibitem{Bertoin}
\leavevmode\vrule height 2pt depth -1.6pt width 23pt, {\em Eternal solutions to
  {S}moluchowski's coagulation equation with additive kernel and their
  probabilistic interpretations}, Ann. Appl. Probab., 12 (2002), pp.~547--564.

\bibitem{Bertoin2}
\leavevmode\vrule height 2pt depth -1.6pt width 23pt, {\em Self-attracting
  {P}oisson clouds in an expanding universe}, Comm. Math. Phys., 232 (2002),
  pp.~59--81.

\bibitem{Bingham}
{\sc N.~H. Bingham, C.~M. Goldie, and J.~L. Teugels}, {\em Regular variation},
  vol.~27 of Encyclopedia of Mathematics and its Applications, Cambridge
  University Press, Cambridge, 1987.

\bibitem{Chandra}
{\sc S.~Chandrasekhar}, {\em Stochastic problems in physics and astronomy},
  Rev. Modern. Phys., 15 (1943), pp.~1--89.

\bibitem{daCosta}
{\sc F.~P. Da~Costa}, {\em On the dynamic scaling behaviour of solutions to the
  discrete {S}moluchowski equations}, Proc. Edinburgh Math. Soc. (2), 39
  (1996), pp.~547--559.

\bibitem{DT}
{\sc M.~Deaconu and E.~Tanr{\'e}}, {\em {S}moluchowski's coagulation equation:
  probabilistic interpretation of solutions for constant, additive and
  multiplicative kernels}, Ann. Scuola Norm. Sup. Pisa Cl. Sci. (4), 29 (2000),
  pp.~549--579.

\bibitem{Derrida}
{\sc B.~Derrida, C.~Godr\'{e}che, and I.~Yekuitieli}, {\em Scale-invariant
  regimes in one-dimensional models of growing and coalescing droplets}, Phys.
  Rev. A, 44 (1991), pp.~6241--6251.

\bibitem{Drake}
{\sc R.~L. Drake}, {\em A general mathematical survey of the coagulation
  equation}, in Topics in Current Aerosol Research, G.~M. Hidy and J.~R. Brock,
  eds., no.~2 in International reviews in Aerosol Physics and Chemistry,
  Pergammon, 1972, pp.~201--376.

\bibitem{Escobedo}
{\sc M.~Escobedo, S.~Mischler, and B.~Perthame}, {\em Gelation in coagulation
  and fragmentation models}, Comm. Math. Phys., 231 (2002), pp.~157--188.

\bibitem{Feller}
{\sc W.~Feller}, {\em An introduction to probability theory and its
  applications. {V}ol. {II}.}, Second edition, John Wiley \& Sons Inc., New
  York, 1971.

\bibitem{Friedlander}
{\sc S.~Friedlander}, {\em Smoke, Dust and Haze: Fundamentals of Aerosol
  Behavior}, Wiley, NY, 1977.

\bibitem{Gallay}
{\sc T.~Gallay and A.~Mielke}, {\em Convergence results for a coarsening model
  using global linearization}, J. Nonlinear. Sci, 13 (2003), pp.~311--346.

\bibitem{Jeon}
{\sc I.~Jeon}, {\em Existence of gelling solutions for
  coagulation-fragmentation equations}, Comm. Math. Phys., 194 (1998),
  pp.~541--567.

\bibitem{Kingman}
{\sc J.~F.~C. Kingman}, {\em Poisson Processes}, vol.~3 of Oxford Studies in
  Probability, The Clarendon Press Oxford University Press, New York, 1993.
\newblock Oxford Science Publications.

\bibitem{Kiorboe}
{\sc T.~Kiorb{\o}e}, {\em Formation and fate of marine snow: small-scale
  processes with large-scale implications}, Scientia Marina, 66 (2001),
  pp.~67--71.

\bibitem{KP}
{\sc M.~Kreer and O.~Penrose}, {\em Proof of dynamical scaling in
  {S}moluchowski's coagulation equation with constant kernel}, J. Statist.
  Phys., 75 (1994), pp.~389--407.

\bibitem{Lee}
{\sc M.~H. Lee}, {\em A survey of numerical solutions to the coagulation
  equation}, J. Phys A:Math. Gen., 34 (2001), pp.~10219--10241.

\bibitem{MPY}
{\sc T.-Y. Lee, G.~Menon, and R.~L. Pego}, {\em Note on thinning of renewal
  processes with heavy tails}.
\newblock In preparation.

\bibitem{Lorentz}
{\sc G.~G. Lorentz}, {\em Bernstein polynomials}, Chelsea Publishing Co., New
  York, second~ed., 1986.

\bibitem{McL}
{\sc J.~B. McLeod}, {\em On an infinite set of non-linear differential
  equations}, Quart. J. Math. Oxford Ser. (2), 13 (1962), pp.~119--128.

\bibitem{MP2}
{\sc G.~Menon and R.~L. Pego}, {\em Dynamical scaling in {S}moluchowski's
  coagulation equations: uniform convergence}.
\newblock Submitted.

\bibitem{NP1}
{\sc B.~Niethammer and R.~L. Pego}, {\em Non-self-similar behavior in the {LSW}
  theory of {O}stwald ripening}, J. Statist. Phys., 95 (1999), pp.~867--902.

\bibitem{Niwa}
{\sc H.~S. Niwa}, {\em School size statistics of fish}, J. Theor. Biol., 195
  (1998), pp.~351--361.

\bibitem{Norris}
{\sc J.~R. Norris}, {\em {S}moluchowski's coagulation equation: uniqueness,
  nonuniqueness and a hydrodynamic limit for the stochastic coalescent}, Ann.
  Appl. Probab., 9 (1999), pp.~78--109.

\bibitem{Pillai}
{\sc R.~N. Pillai}, {\em On {M}ittag-{L}effler functions and related
  distributions}, Ann. Inst. Statist. Math., 42 (1990), pp.~157--161.

\bibitem{Silk}
{\sc J.~Silk and S.~D. White}, {\em The development of structure in the
  expanding universe}, Astrophysical J., 223 (1978), pp.~L59--L62.

\bibitem{Tsoukatos}
{\sc K.~P. Tsoukatos}, {\em Heavy and light traffic regimes for $M|G|_\infty$
  traffic models}, PhD thesis, University of Maryland, College Park, Electrical
  Engineering Dept., College Park, MD 20742, May 1999.

\bibitem{DE1}
{\sc P.~G.~J. van Dongen and M.~H. Ernst}, {\em Scaling solutions of
  {S}moluchowski's coagulation equation}, J. Statist. Phys., 50 (1988),
  pp.~295--329.

\bibitem{Ziff}
{\sc R.~M. Ziff}, {\em Kinetics of polymerization}, J. Statist. Phys., 23
  (1980), pp.~241--263.

\end{thebibliography}
\end{document}